\documentclass{iopjournal}

\usepackage{amsmath,amssymb,amsfonts}
\usepackage{graphicx}
\usepackage{dsfont}
\usepackage{cancel}
\usepackage{xcolor}

\newcommand{\ii}{{\rm i}}
\newcommand{\dd}{{\rm d}}
\newcommand{\ee}{{\rm e}}

\begin{document}

\articletype{Paper}

\title{Landau levels in a time-dependent magnetic field: the Madelung fluid perspective}

\author{Nicolas Perez$^{1,2,*}$\orcid{0000-0001-5067-102X} and Eyal Heifetz$^{1}$\orcid{0000-0002-3584-3978}}

\affil{$^1$Department of Geophysics, Tel Aviv University, 69978 Tel Aviv, Israel}

\affil{$^2$The Steinhardt Museum of Natural History, Tel Aviv University, 12 Klausner Street, 6901127 Tel Aviv, Israel}

\affil{$^*$Author to whom any correspondence should be addressed.}

\email{nicolasperez@tauex.tau.ac.il}


\begin{abstract}

We revisit the quantum dynamics of a charged particle in a time-dependent magnetic field, a fundamental problem exhibiting rich non-adiabatic behaviour, from the complementary perspective of the Madelung fluid formulation. We first analyse the system within standard quantum mechanics using perturbation theory around the Landau levels, and then address the same problem through the Madelung perspective. We show that the hydrodynamic formulation not only yields an intuitive derivation of the exact solution, it also provides a clear physical interpretation of non-adiabatic quantum evolution in terms of mechanical energy transfers. In this picture, the sloshing oscillations of the wave function arise from deviations from the force balance between the magnetic Lorentz force and the gradient of the Bohm potential within the Landau levels. More broadly, our study illustrates how the Madelung approach reveals unexpected analogies between quantum dynamics and phenomena familiar from geophysical fluid dynamics.

\end{abstract}

\section{Introduction}

\begin{figure}[t]
\begin{center}
\includegraphics[scale=0.45]{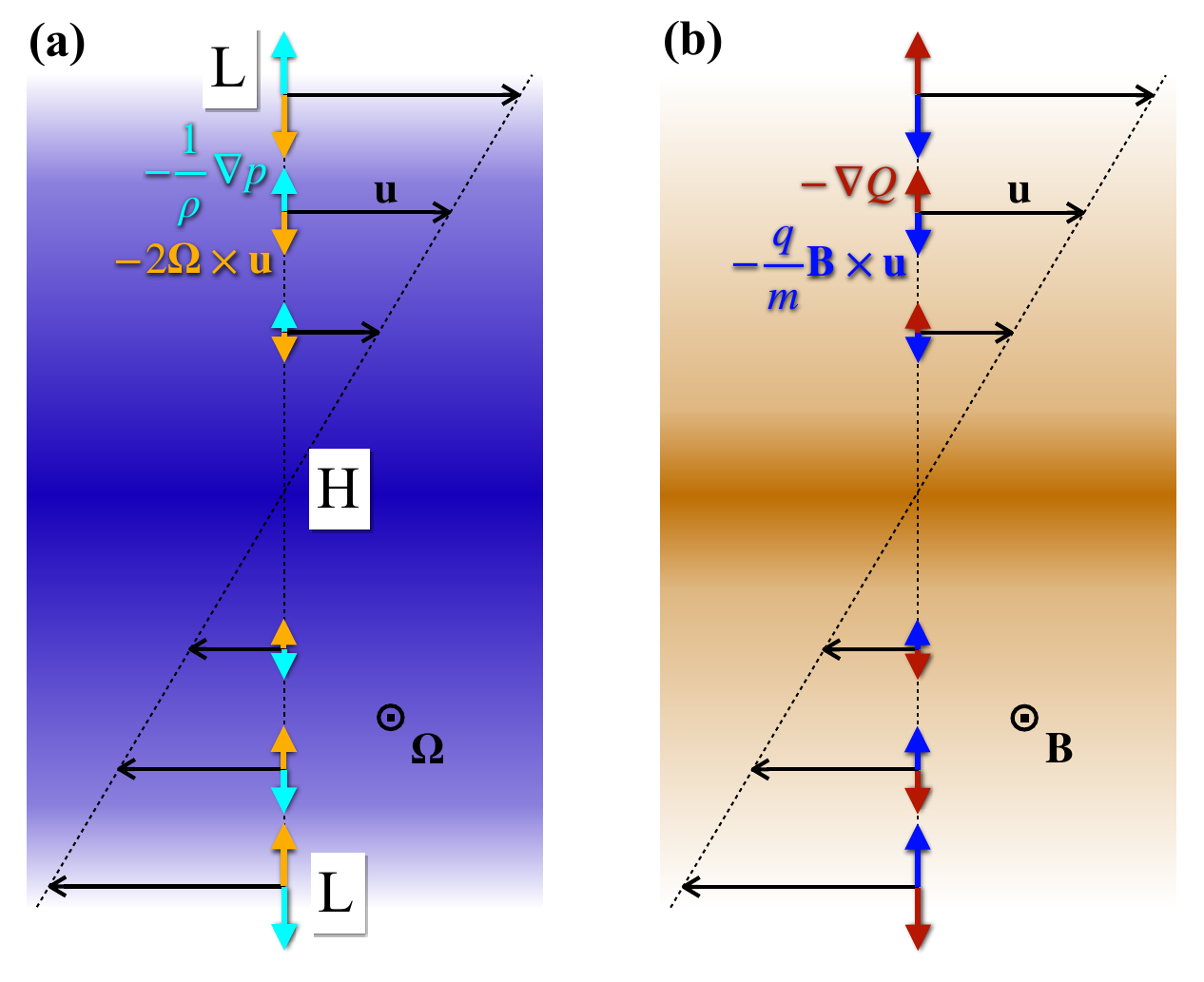}
\end{center}
\caption{\label{fig:balance} \textbf{(a)} Geostrophic balance of a two-dimensional shear flow. The stationary velocity field ${\bf u}$ is of a plane Couette shear flow such that the pressure gradient force $-\frac{1}{\rho} \boldsymbol{\nabla} p$ and the Coriolis force $-2\boldsymbol{\Omega}\times{\bf u}$, are exactly in balance everywhere. $p$ denotes pressure (where $H$ and $L$ correspond respectively to regions of high and low pressure) and $\rho$ denotes the fluid density. The flow is embedded within a rotating system with angular velocity $\boldsymbol{\Omega}$, pointing upward, perpendicular to the shear plane. \textbf{(b)} Landau level geostrophic-like balance analogue. A two-dimensional Madelung fluid representing a configuration of a charged particle's wave function, under a transverse magnetic field ${\bf B}$, pointing upward. The gradient of the Bohm potential $Q$ plays an equivalent role of the pressure gradient force and is in balance with the Lorentz force $-\frac{q}{m}{\bf B}\times{\bf u}$ (where $q$ and $m$ are the particle's electric charge and mass, respectively).}
\end{figure}

Less than a year after Schrödinger proposed his eponymous equation, Madelung established an alternative formulation of it as a set of partial-differential equations describing a fluid \cite{madelung1927quantum}, whose density and current are connected to the modulus of the wave function and the gradient of its phase. This conceptual fluid, known today as the Madelung fluid, was later used by Bohm \cite{bohm1952suggested} to revive and develop the pilot wave interpretation of quantum mechanics, formulated originally by de Broglie. Over the last few decades, Madelung hydrodynamics has been widely used in the study of Bose–Einstein condensates \cite{dalfovo1999theory,pethick2008bose,chavanis2011mass,carles2012madelung,andreev2014dispersion,andreev2021quantum}, where it provides a fluid-dynamical reinterpretation of the Gross–Pitaevskii equation and a transparent description of superfluid phenomena. This approach has proved useful for analysing collective dynamics and hydrodynamic limits of condensates, and for connecting quantum mean-field models with classical continuous descriptions. Besides, this formalism has regained attention in the recent years \cite{berry2023time,berry2024kinetically}, and it has been exploited for the purpose of bringing new insights on quantum problems based on their analogies with fluid dynamics \cite{heifetz2015toward,heifetz2016entropy,heifetz2020madelung,heifetz2021zero,heifetz2023broglie}. In this perspective, Heifetz \textit{et al} \cite{heifetz2021zero} demonstrated that the two-dimensional Madelung hydrodynamical equations associated to the Schrödinger equation of a spinless charged particle in a transverse magnetic field are those of a compressible flow in a rotating frame, with zero absolute vorticity. They mapped the stationary solutions of the Madelung equations to the Landau levels of the charged particle, thus providing a mechanistic interpretation of these eigenstates as steady plane Couette shear flows, in equilibrium under balanced forces acting perpendicularly to the flow: the gradient of the quantum (Bohm) potential, which acts from the centre of the shear flow outward, and the Lorentz magnetic force, which acts inward. This is analogous to the geostrophic balance in high pressure systems, in the atmosphere and the ocean: the pressure gradient force acts outwards, whereas the Coriolis force acts inwards (Figure \ref{fig:balance}).

In this paper, we examine the non-stationary dynamics of the Madelung fluid when the flow is set out of balance. One way to do this is to consider that the initial wave function is an eigenstate of the Schrödinger equation -- i.e. that the corresponding Madelung flow is stationary -- and change the magnetic field with time, the question being that of the subsequent dynamics of the wave function and the corresponding Madelung flow. This particular problem has been addressed in different works with the perspective of emphasising specific dynamical properties of the Landau levels under time-varying parameters. It thus naturally relates to the quantum adiabatic theorem \cite{born1928beweis,berry1984quantal}, which has triggered a lot of interest owing to the great importance of geometric phases in driven condensed matter systems \cite{berry1984quantal,xiao2010berry}. For instance, the study of \cite{chee2009landau} exhibits geometric phases in Landau level dynamics with time-dependent electric fields under adiabatic conditions, while \cite{kim2014landau} focuses on interlevel transitions with a time-dependent transverse magnetic field under adiabatic or non-adiabatic conditions. To our knowledge, the only relatively recent work that proposed a hydrodynamical perspective to this problem -- although not with the Madelung transformation -- is \cite{sutherland1998exact}. It deals with a different problem of one-dimensional $N$-body systems with inverse-square-potential pair interactions and a time-varying harmonic trap, but only addresses the dynamics of its coherent states. The analysis reveals a sloshing behaviour of these coherent states, and proposes to associate it to persistent oscillations in one-dimensional superfluids.

We address the problem of Landau levels dynamics under a time-varying transverse magnetic field with a fluid perspective, through the Madelung formulation. However, for the sake of comparison and familiarity, we first present the problem in a quantum mechanics fashion, similar to \cite{kim2014landau}, which allows us to obtain perturbatively an approximate expression of the wave function through the so-called squeezing operator. Next, we introduce the hydrodynamical formulation of the same problem, whose stationary solutions are plane Couette shear flows. We show that this formulation provides a mechanistic interpretation of the Landau level dynamics, and naturally yields the exact solution, valid for all Landau levels, that was originally obtained by \cite{lewis1969exact} and is consistent with the sloshing behaviour discussed by \cite{sutherland1998exact}. We discuss further the relation between this dynamics and the phenomenon of geostrophic adjustment in geophysical fluid dynamics, that we can in turn relate to non-adiabatic quantum evolution. We then derive the energy constant of motion of the linearised dynamics of small perturbations (denoted in geophysical fluid dynamics as pseudo-energy) whose positive-definiteness reveals both stability of the perturbations and the possibility for hysteresis behaviour.

Although the quantum problem addressed in this paper can be obtained via other methods, this paper aims to demonstrate that the Madelung framework provides a straightforward semi-classical description of the wave function, allowing one to interpret quantum evolution in a physically intuitive way, by means of a local fluid velocity and quantum enthalpy connected to each other through conservation laws. Besides, this complementary description provides an intuitive derivation of the exact mathematical solution for the time-dependent Landau levels, in contrast with more conventional approaches such as perturbation theory, which do not unravel the local structure of the wave function. Additionally, this work aims to frame non-adiabatic quantum evolution with the Madelung fluid perspective, by establishing a parallel with ageostrophic dynamics in geophysical fluids.

\section{Quantum mechanics perspective} \label{sec:quantum}

In this section, we first revisit the problem of the dynamics of a two-dimensional charged particle in a transverse homogeneous magnetic field, and recall a basis of eigenstates of the corresponding stationary Schrödinger equation, the celebrated Landau levels. Then, building upon \cite{kim2014landau}, we propose to address the non-stationary case -- with a time-dependent magnetic field -- by expanding the wave function in a time-dependent basis, which allows one to separate the adiabatic and non-adiabatic contributions of the evolution. Finally, we discuss the conclusions of this analysis regarding the quantum adiabatic theorem \cite{born1928beweis,berry1984quantal}.

\subsection{Definition of the quantum problem, Landau levels} \label{part:Landau}

We consider a non-relativistic, spinless particle of mass $m$ and charge $q$, evolving on the unbounded two-dimensional plane $(x,y)$ and subjected to an external electromagnetic field $\left( \mathbf{E} , \mathbf{B} \right)$. Its wave function $\psi$ is described by the Schrödinger equation
\begin{equation} \label{eq:Schrodinger}
    \begin{split}
        &\ii \hbar \partial_t \psi = \frac{\left( -\ii \hbar \boldsymbol{\nabla} - q \mathbf{A} \right)^2}{2 m} \psi + q V \psi \ , \\[6pt]
        \text{with} \quad &\mathbf{B} = \boldsymbol{\nabla} \times \mathbf{A} \quad \text{and} \quad \mathbf{E} = -\boldsymbol{\nabla} V - \partial_t \mathbf{A} \ ,
    \end{split}
\end{equation}
where $\mathbf{B} = B(t) \mathbf{e}_z$ is a homogeneous transverse magnetic field, and $\mathbf{E}$ an induced electric field. Following \cite{heifetz2021zero}, equation \eqref{eq:Schrodinger} can be simplified in the Landau gauge\footnote{Generally speaking, the choice \eqref{eq:Landau_gauge} is valid regarding the Maxwell equations as long as the typical variation time of $B(t)$ remains large compared to the time it takes electromagnetic waves to travel through the system (quasi-static approximation), which is assumed here. Otherwise, $\mathbf{E}$ and $\mathbf{B}$ together do not obey the induction equation.}:
\begin{equation} \label{eq:Landau_gauge}
    \mathbf{A} = ( A_0 - B(t) y ) \mathbf{e}_x \quad \text{and} \quad V = V_0 \ .
\end{equation}

In this gauge, equation \eqref{eq:Schrodinger} is invariant in the $x$ direction (hereafter, this coordinate will be referred to as zonal, adopting the geophysical fluid dynamics jargon in the equivalent context\footnote{In this paper we aim to connect the quantum mechanics and the geophysical fluid dynamics perspectives when addressing this problem, thus we use familiar jargons from the two disciplines.}), thus we can consider Fourier modes of wave number $k$ in this direction:
\begin{equation} \label{eq:Fourier_mode_x}
    \psi(x,y,t) = \ee^{\ii k x} \hat{\psi} (y,t) \ .
\end{equation}

For a given wave number $k$, choosing $A_0 = \hbar k / q$ and $V_0 = 0$ further reduces equation \eqref{eq:Schrodinger} to
\begin{equation} \label{eq:Schrodinger_bis_dimensions}
    \ii \partial_t \hat{\psi} = -\frac{\hbar^2}{2 m} \partial_{yy} \hat{\psi} + \frac{q^2 B(t)^2}{2 m} y^2 \hat{\psi} \ .
\end{equation}

Let us note that, for any alternative gauge $(\mathbf{A}',V') = \left( \mathbf{A} + \frac{\hbar}{q} \boldsymbol{\nabla} \chi , V - \frac{\hbar}{q} \partial_t \chi \right)$, a solution of \eqref{eq:Schrodinger} is given by $\psi' = \exp \left( \ii \chi \right) \psi$, which only differs by the gauge phase factor $\chi(x,y,t)$\footnote{The gauge choice for the electromagnetic potentials has no influence on the dynamics of the Madelung fluid since it does not affect the hydrodynamical fields, see the definitions \eqref{eq:fields} in section \ref{sec:fluid}.}. From now on, we consider this 1D form and drop the $\hat{}$ on the function $\hat{\psi}(y,t)$. Besides, let us use dimensionless quantities, obtained by dividing time $t$ by a time scale $T$ and the spatial coordinates $x,y$ by a length $L$, chosen such that $L^2 / T = \hbar / m$. We also use the dimensionless magnetic field $b = (qT/m) B$ (which we will always assume positive from now on)\footnote{In the case of a constant magnetic field, if $L^2 / T = \hbar / m$ and $b=1$, then $T$ is the inverse of the cyclotron frequency and $L$ is called the magnetic length.}, so that equation \eqref{eq:Schrodinger_bis_dimensions} becomes
\begin{equation} \label{eq:Schrodinger_bis}
    \ii \partial_t \psi = -\frac{1}{2} \partial_{yy} \psi + b(t)^2 \frac{y^2}{2} \psi = \mathcal{H} \psi \ .
\end{equation}

The problem is now entirely defined by the initial condition and the function $b(t)$. In the case of a constant magnetic field $b$, the stationary solutions of equation \eqref{eq:Schrodinger_bis} read as
\begin{equation} \label{eq:Fourier_mode}
    \psi(y,t) = \ee^{-\ii \omega t} \Psi (y) \ ,
\end{equation}
for which equation \eqref{eq:Schrodinger_bis} becomes a second-order ordinary differential equation in the $y$ coordinate (hereafter, this coordinate will be referred to as meridional):
\begin{equation} \label{eq:harmonic_oscillator}
    \omega \Psi = -\frac{1}{2} \frac{\dd^2 \Psi}{\dd y^2} + \frac{1}{2} b^2 y^2 \Psi \ ,
\end{equation}
which is the well-known eigenvalue equation of a 1D quantum harmonic oscillator. Since the meridional direction is unbounded, the physically acceptable solutions of \eqref{eq:harmonic_oscillator} are the Hermite-Gauss functions
\begin{equation} \label{eq:Landau}
    \Psi_{n,b} (y) = \frac{1}{\sqrt{2^n \, n!}} \left( \frac{b}{\pi} \right)^\frac{1}{4} \ee^{-\frac{b}{2} y^2} H_{n} \left( \sqrt{b} y \right) \; \text{and} \; \omega_{n,b} = b \left( n + \frac{1}{2} \right) \; \text{with} \; n \in \mathbb{N} \ .
\end{equation}

These are known as the Landau levels, which are the keystone of quantum evolution under magnetic fields. In the rest of the paper, we examine how these eigenstates behave following a change of magnetic field. The response of Landau levels -- and more generally quantum states of charged particles -- to time-dependent magnetic fields has been examined in several contexts, including interlevel transitions, adiabaticity, wave-packet evolution and lattice-based quantum systems \cite{berry1984quantal,breuer1989quantum,sundaram1999wave,chee2009landau,ardenghi2013landau,kim2014landau,razzoli2020continuous,morais2024landau}. These studies have clarified the quantum evolution of the wave function under magnetic forcing, revealing breathing and oscillatory behaviours. However, the physical origin of these motions is usually discussed in terms of quantum transitions (as presented hereafter in \ref{part:squeezing}) and interference phenomena. The present work adopts a complementary viewpoint based on the Madelung fluid formulation, which we introduce in section \ref{sec:fluid}. By recasting the dynamics in terms of density and velocity fields, it becomes possible to identify the force balance responsible for stationary Landau states and to interpret departures from adiabaticity as mechanical-energy exchanges within an effective quantum fluid. To emphasise our contribution, we start in \ref{part:squeezing} by addressing the problem with a perturbative method designed to characterize the dynamics in terms of interlevel transitions.

\subsection{Perturbative expression of the wave function, squeezing and interlevel transitions} \label{part:squeezing}

Let us now describe the evolution of the wave function $\psi$, starting in a Landau level $n_0$ and letting $b$ to vary in time. In this part, we adopt a perturbative approach based on projecting the wave function on a basis of time-dependent Landau states. The evolution equation is \eqref{eq:Schrodinger_bis}, with the initial condition
\begin{equation} \label{eq:IC_wave_function}
    \psi (y,0) = \Psi_{n_0 , b_0} (y) \ ,
\end{equation}
where the functions $\Psi_{n,b}$ are defined by expression \eqref{eq:Landau} and $b_0 = b(0)$. If $b$ varies very slowly, one can intuitively predict that the wave function $\psi (y,t)$ remains close to $\Psi_{n_0 , b(t)} (y)$ at time $t > 0$ (see the comments regarding the adiabatic theorem in \ref{part:adiabatic}). To quantify this, we consider the projection of the wave function on the time-dependent orthonormal basis $\{ \Psi_{m,b(t)} (y) \, | \, m \in \mathbb{N} \}$:
\begin{equation} \label{eq:Berry_projections}
    \psi(y,t) = \sum_{m \geq 0} \phi_m(t) \Psi_{m,b(t)} (y) \ ,
\end{equation}
which is directly inspired by the analysis of \cite{berry1984quantal,kim2014landau}. The normalization of the wave function implies that the sum of $|\phi_m (t)|^2$ over all positive integers $m$ is equal to 1, and we also have $\phi_m (0) = \delta_{m n_0}$ by definition of the initial state. This decomposition \eqref{eq:Berry_projections} amounts to writing the evolution of the wave function in a frame that follows the variations of $b$. In the rest of this section, it will be convenient to adopt a bra-ket notation for the functions of $y$: we represent the function $\psi(y,t)$ with the ket $| \psi \rangle$. This notation will be particularly convenient to use the Hermitian product defined as
\begin{equation}
    \langle \psi_1 | \psi_2 \rangle = \int_{-\infty}^{+\infty} \psi_{1}^* (y) \psi_2 (y) \, \dd y \ .
\end{equation}

We will also denote the eigenfunctions $\Psi_{m,b}$ and eigenvalues $\omega_{m,b}$ simply by $| m \rangle$ and $\omega_m$ respectively, but always keeping in mind that they are time-dependent through $b(t)$. Injecting expression \eqref{eq:Berry_projections} into equation \eqref{eq:Schrodinger_bis} and projecting on the basis function of a given level $n$, one gets
\begin{equation} \label{eq:first_projection}
    \ii \dot{\phi}_n = \omega_n \phi_n - \ii \dot{b} \sum_{m \geq 0} \langle n | \partial_b | m \rangle \, \phi_m \ ,
\end{equation}
where we note the time-differentiation $\dd A / \dd t = \dot{A}$. The sum in the RHS of equation \eqref{eq:first_projection} contains only interlevel terms ($m \neq n$). To see that, we recall that the functions $\Psi_{n,b}$ are real-valued and normalised, therefore
\begin{equation} \label{eq:zero_diagonal}
    \langle n | \partial_b | n \rangle = \int_{-\infty}^{+\infty} \Psi_{n,b} (y) \partial_b \Psi_{n,b} (y) \, \dd y = \frac{1}{2} \frac{\partial}{\partial b} \int_{-\infty}^{+\infty} \Psi_{n,b}^2 (y) \, \dd y = 0 \ .
\end{equation}

For the terms $m \neq n$ in the sum, we can show that\footnote{This is equation (8) in \cite{berry1984quantal}.}
\begin{equation}
    \langle n | \partial_b | m \rangle = \frac{\langle n | \partial_b \mathcal{H} | m \rangle}{\omega_m - \omega_n} \ ,
\end{equation}
where $\mathcal{H}$ is the differential operator in the RHS of equation \eqref{eq:Schrodinger_bis}. Therefore, $\partial_b \mathcal{H}$ is equal to the operator $b y^2$. Using the formulation of $\mathcal{H}$ with time-dependent ladder operators\footnote{These are such that $\mathcal{H} = b \left( c^\dagger c + \frac{1}{2} \right)$, where $^\dagger$ indicates the adjoint. We have $c = y\sqrt{b/2} + \partial_y / \sqrt{2b}$ and $c^\dagger = y\sqrt{b/2} - \partial_y / \sqrt{2b}$. They act on the eigenfunctions as $c | m \rangle = \sqrt{m} \: | m-1 \rangle$ and $c^\dagger | m \rangle = \sqrt{m+1} \: | m+1 \rangle$.} $c$ and $c^\dagger$, we get
\begin{equation} \label{eq:complicated_stuff}
    \partial_b \mathcal{H} = \frac{1}{2} \left( c + c^\dagger \right)^2 \ ,
\end{equation}
which yields the following formulation of equation \eqref{eq:first_projection}:
\begin{equation} \label{eq:recurrence_Landau}
    \ii \dot{\phi_n} = \omega_n \phi_n - \ii \frac{\dot{b}}{4 b} \left( \sqrt{(n+2)(n+1)} \phi_{n+2} - \sqrt{n(n-1)} \phi_{n-2} \right) \ .
\end{equation}

\begin{figure}[t]
\begin{center}
\includegraphics[scale=0.35]{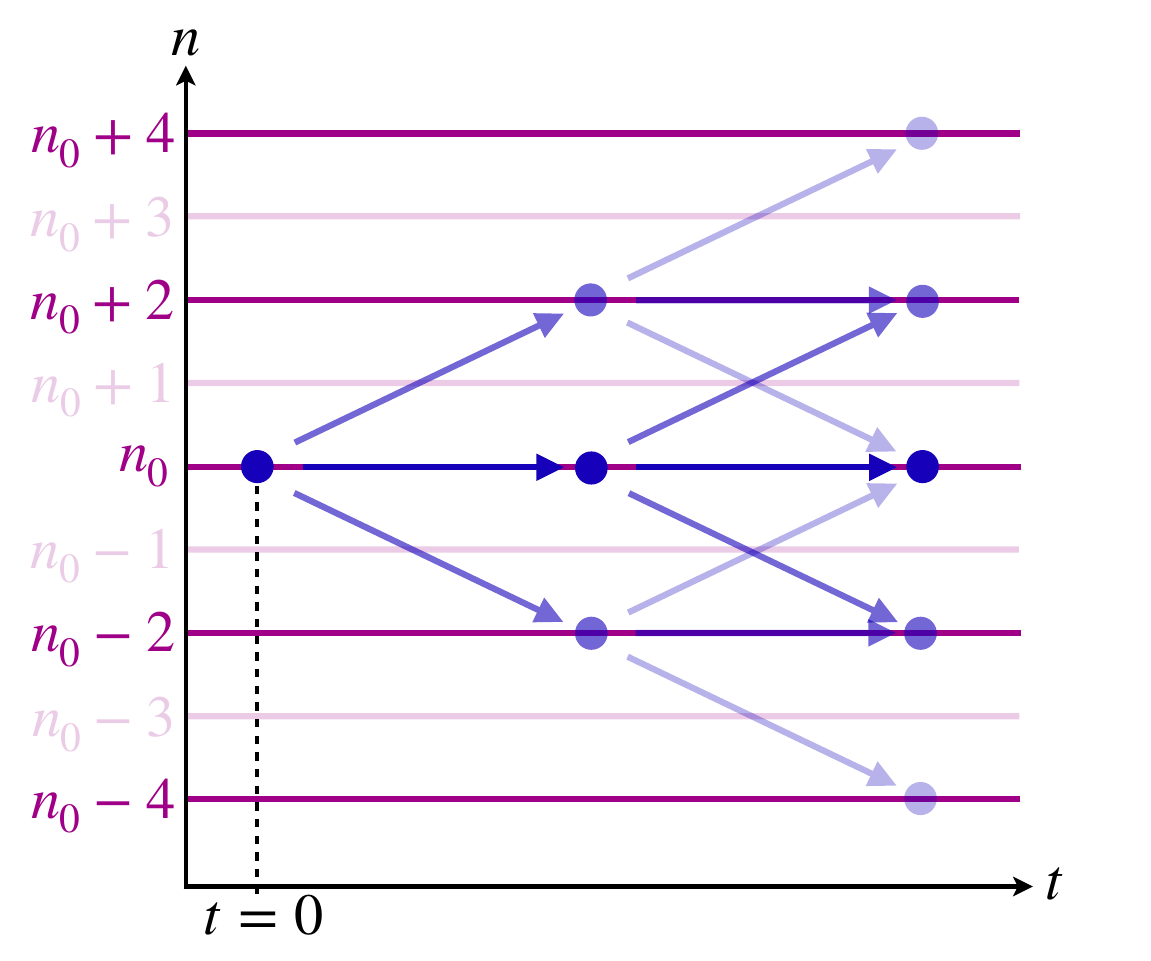}
\end{center}
\caption{\label{fig:interlevel_transitions} Illustration of the interlevel transitions embodied by the coupled differential equations \eqref{eq:recurrence_Landau}. If the wave function is initially in the level $|n_0 \rangle$, it then successively populates the (time-dependent) levels $|n_0 \pm 2 \rangle$, $|n_0 \pm 4 \rangle$, and so on as $b(t)$ varies.}
\end{figure}

The recurrence relation \eqref{eq:recurrence_Landau} can be recast into a single equation, introducing the vector of the projection coefficients, $\overrightarrow{\phi} = \begin{pmatrix} \phi_0 & \phi_1 & \phi_2 & ... \, \end{pmatrix}^\top = \sum_{m \geq 0} \phi_m (t) \overrightarrow{e_m}$:
\begin{equation} \label{eq:evolution_Heisenberg}
    \ii \frac{\dd}{\dd t} \overrightarrow{\phi} = b(t) \left( C^\dagger C + \frac{1}{2} \right) \overrightarrow{\phi} + \ii \frac{\dot{b}(t)}{4 b(t)} \left( C^{\dagger 2} - C^2 \right) \overrightarrow{\phi} \ ,
\end{equation}
where the ladder operators $C$ and $C^\dagger$ are defined such that
\begin{equation}
    C \overrightarrow{e_m} = \sqrt{m} \: \overrightarrow{e_{m-1}} \quad \text{and} \quad C^\dagger \overrightarrow{e_m} = \sqrt{m+1} \: \overrightarrow{e_{m+1}} \ .
\end{equation}

Several comments can be made about expression \eqref{eq:evolution_Heisenberg}:

\begin{itemize}
    \item First of all, it is a Schrödinger equation in a 
    space that represents the time evolution of the basis of eigenstates. The first term in the RHS of \eqref{eq:evolution_Heisenberg} represents the adiabatic contribution, i.e. the dynamical part that consists in the Landau levels simply following the changes of the magnetic field $b$. Without the second term, any eigenstate at $t=0$ remains an eigenstate (in the same level $n_0$) at time $t>0$.
    
    \item The second term in the RHS -- which was alternatively obtained in \cite{kim2014landau} with a Klein-Gordon formulation of the problem -- represents the non-adiabatic part of the evolution, which quantifies the wave function departing from the expected Landau level $\Psi_{n_0 , b(t)}$. This term produces interlevel transitions between a level $m$ and the levels $m \pm 2$ (Figure \ref{fig:interlevel_transitions})\footnote{The levels $m \pm 1$ do not interact with the level $m$ during the evolution as a result of parity symmetry being preserved by varying $b$ with time: an eigenstate of level $m$ has the same parity as $m$ in the $y$ direction, which is preserved with time.}.
    
    \item Finally, the fact that the term in $\dot{b}$ is non-diagonal is symptomatic of the absence of geometric or Berry phase during the evolution process. This comes from \eqref{eq:zero_diagonal}, which is characteristic of the absence of phase in the eigenstates and the one-dimensional nature of the problem. If the problem is regarded in higher dimension or with more varying parameters (for instance an electric field or another component for the magnetic field), then a geometric phase may arise in the evolution, as in \cite{chee2009landau}.
\end{itemize}

In order to focus on the effect of the second term in equation \eqref{eq:evolution_Heisenberg}, we will adopt the interaction picture (see for instance \cite{bruus2004many}) and define the vector
\begin{equation} \label{eq:interaction_picture}
    \overrightarrow{\Phi} = \exp \left\{ \ii \left( C^\dagger C + \frac{1}{2} \right) \Theta(t) \right\} \overrightarrow{\phi} = \sum_{m \geq 0} \ee^{\ii \left( m + \frac{1}{2} \right) \Theta(t)} \phi_m (t) \overrightarrow{e_m} \ ,
\end{equation}
where we have noted $\Theta(t) = \int_{0}^{t} b(t_1) \, \dd t_1$. The transformation \eqref{eq:interaction_picture} separates the dominant dynamical phases of the wave functions -- i.e. the terms $\exp \left\{ -\ii \left( m + 1/2 \right) \Theta(t) \right\}$ that describe the respective adiabatic evolution of each coefficient $\phi_m$ -- from the interlevel evolution, which underlies the non-adiabatic dynamics. Inserting \eqref{eq:interaction_picture} into \eqref{eq:evolution_Heisenberg} yields
\begin{equation}
    \begin{split}
        \frac{\dd}{\dd t} \overrightarrow{\Phi} &= \frac{\dot{b}}{4 b} \exp \left\{ \ii \left( C^\dagger C + \frac{1}{2} \right) \Theta(t) \right\} \left( C^{\dagger 2} - C^2 \right) \exp \left\{ -\ii \left( C^\dagger C + \frac{1}{2} \right) \Theta(t) \right\} \overrightarrow{\Phi} \\[6pt]
        &= \frac{\dot{b}}{4 b} \left( \ee^{2 \ii \Theta(t)} C^{\dagger 2} - \ee^{-2 \ii \Theta(t)} C^2 \right) \overrightarrow{\Phi} = \mathcal{A}(t) \overrightarrow{\Phi} \ .
    \end{split}
\end{equation}

Therefore, we can express the vector $\overrightarrow{\Phi}$ at time $t > 0$ as
\begin{equation} \label{eq:time-ordered_exp}
    \begin{split}
        \overrightarrow{\Phi} (t) &= \mathcal{S}(t) \overrightarrow{\Phi} (0) = \mathcal{S}(t) \overrightarrow{e_n} \ , \quad \text{with the evolution operator} \\[6pt]
        \mathcal{S}(t) &= 1 + \int_{0}^t \dd t_1 \, \mathcal{A}(t_1) + \int_{0}^t \dd t_1 \, \int_{0}^{t_1} \dd t_2 \, \mathcal{A}(t_1) \mathcal{A}(t_2) + \int_{0}^{t} \dd t_1 \, \int_{0}^{t_1} \dd t_2 \, \int_{0}^{t_2} \dd t_3 \, \mathcal{A}(t_1) \mathcal{A}(t_2) \mathcal{A}(t_3) + ... \\[6pt]
        &= \mathcal{T} \left\{ \exp \left( \int_{0}^t \dd t' \, \mathcal{A}(t') \right) \right\} \ .
    \end{split}
\end{equation}

The final expression in \eqref{eq:time-ordered_exp} is a time-ordered\footnote{The time-ordered exponential differs from the natural exponential since the operators $\mathcal{A}(t_1)$ and $\mathcal{A}(t_2)$ do not commute for $t_1 \neq t_2$ \cite{bruus2004many}.} version of the so-called squeeze operator, which is a well-known evolution operator in the fields of quantum optics and quantum information. Although there is no exact expression for $\mathcal{S}(t)$, as far as we know, it provides a natural perturbation formula for the non-adiabatic effects, as long as the integral of $\mathcal{A}(t)$ remains very small, i.e. as $b$ is varied very slowly, close to adiabaticity. Keeping only the first-order term in the expression in the second line of \eqref{eq:time-ordered_exp}, we have
\begin{equation} \label{eq:first-order_evolution}
    \begin{split}
        \overrightarrow{\Phi} (t) \approx \overrightarrow{e_{n_0}} &+ \zeta(t) \sqrt{(n_0 +2)(n_0 +1)} \: \overrightarrow{e_{n_0 +2}} - \zeta(t)^* \sqrt{n_0 (n_0 -1)} \: \overrightarrow{e_{n_0 -2}} \ , \\[6pt]
        &\text{with} \quad \zeta(t) = \int_{0}^t \dd t' \, \frac{\dot{b}(t')}{4 b(t')} \ee^{2 \ii \Theta(t')} \ .
    \end{split}
\end{equation}

Expression \eqref{eq:first-order_evolution} characterises the rate at which interlevel transitions occur, i.e. at which the neighbouring levels $n_0 \pm 2$ are populated as a result of non-adiabaticity. As for the wave function, we thus obtain
\begin{equation} \label{eq:first-order-adiabatic}
    \begin{split}
        \psi(y,t) \approx \; &\ee^{-\ii \left( n_0 + \frac{1}{2} \right) \Theta(t)} \: \Psi_{n_0 ,b(t)} (y) \\[6pt]
        + \: &\ee^{-\ii \left( n_0 + 2 + \frac{1}{2} \right) \Theta(t)} \: \zeta(t) \: \sqrt{(n_0 +2)(n_0 +1)} \: \Psi_{n_0 +2,b(t)} (y) \\[6pt]
        - \: &\ee^{-\ii \left( n_0 - 2 + \frac{1}{2} \right) \Theta(t)} \: \zeta(t)^* \: \sqrt{n_0 (n_0 -1)} \: \Psi_{n_0 -2,b(t)} (y) \ .
    \end{split}
\end{equation}

This approximate solution only takes into account the nearest interlevel transition terms ($n_0 \pm 2$). In spite of this approximation, the amplitude factor $\zeta (t)$ remains difficult to express in general. Nevertheless, let us look at the asymptotical limit, assuming that $b$ converges to a final value $b_1$ when $t \rightarrow +\infty$. Since $\zeta$ is bounded and $\dot{\zeta}(t) \rightarrow 0$ when $t \rightarrow +\infty$, the function $\zeta$ converges to a constant value $\zeta_1$, which is small if $b$ varies slowly\footnote{Indeed, assuming that $b(t)$ is monotonic without losing generality, we have $|\zeta(t)| < \int_0^t \dd t' \, |\dot{b}(t') / 4 b(t')| = |\ln ( b_1 / b_0 )|/4$ when $t \rightarrow +\infty$, which is finite.}. Therefore, all the coefficients $\phi_n (t)$ of the wave function converge to a constant amplitude times the dynamical phase factor\footnote{This result can be extrapolated to the full wave function. Expanding the evolution operator $\mathcal{S}(t)$ to all orders, one can show that it converges to a constant operator at infinite time.} $\ee^{-\ii (n + 1/2 ) \Theta(t)}$. Besides, the function $\Theta(t)$ is asymptotically equal to a constant $\Theta_1$ plus the linear function $b_1 t$. Therefore, as $t \rightarrow +\infty$, the modulus of the approximate wave function \eqref{eq:first-order-adiabatic} reads as
\begin{equation}
    \begin{split}
        |\psi| \approx |\Psi_{n_0 ,b_1} (y) &+ \ee^{-2\ii \left( \Theta_1 + b_1 t \right)} \zeta_1 \sqrt{(n_0 +2)(n_0 +1)} \Psi_{n_0 +2,b_1} (y) \\[6pt]
        &- \ee^{2\ii \left( \Theta_1 + b_1 t \right)} \zeta_1^* \sqrt{n_0 (n_0 -1)} \Psi_{n_0 -2,b_1} (y)| \ .
    \end{split}
\end{equation}

Thus, if the magnetic field $b$ converges to a final value $b_1$, the modulus of the wave function eventually oscillates periodically with frequency $2 b_1$ around the adjusted Landau level's modulus $|\Psi_{n_0 , b_1} (y)|^2$, and these oscillations do not fade. However, the expansion \eqref{eq:first-order-adiabatic} is generally not valid at long time, as there is no reason to neglect the next-order integrals in expression \eqref{eq:time-ordered_exp}, which also yield contributions to the levels $n_0 , n_0 \pm 2$ (see Figure \ref{fig:interlevel_transitions}). Nevertheless, the periodic oscillation of frequency $2 b_1$ is a general result which remains true with the full expansion of the evolution operator, since the phase difference between every interlevel term and the main coefficient is a multiple of $2 \Theta(t)$. The final oscillating dynamics is thus the result of both symmetry being preserved and the Landau levels involved in the evolution being separated by frequency gaps that are all multiples of $2 b_1$. However, not only this perturbative approach breaks down at long times, but, even when valid, it does not readily provide either the expression of $\zeta_1$ or the amplitude of the final oscillations. In the next section, we show how the Madelung perspective, applied to the problem \eqref{eq:Schrodinger_bis}-\eqref{eq:IC_wave_function}, naturally yields a solution that cannot be captured by the perturbative approach adopted in this section.

\subsection{Remarks on the adiabatic theorem} \label{part:adiabatic}

The previous analysis must be addressed in light of the adiabatic theorem in quantum mechanics \cite{born1928beweis,berry1984quantal}: generally speaking, if a Hamiltonian $H(t)$ varies slowly (adiabatically) and its discrete eigenvalues $\{ \lambda_n (t) \}$ remain sufficiently separated, then a quantum system initially prepared in the eigenstate $n$ remains in the same state (whose eigenvalue $\lambda_n$ is a function of time) \cite{berry1984quantal}. Such evolution of the wave function is adiabatic in the sense that the information on the state is preserved in time. A requirement for adiabaticity is that the typical time of variation of $H(t)$ must remain large compared to $\hbar/\Delta_n (t)$, where $\Delta_n (t)$ is the energy gap that separates $\lambda_n (t)$ from the closest energy level of $H(t)$. Expression \eqref{eq:first-order-adiabatic} illustrates the limitation of this theorem: the leading-order term of the wave function corresponds to the Landau level adjusted to the value $b(t)$, and the corrective interlevel terms at first order are proportional to $\zeta$, whose amplitude vanishes in the adiabatic limit, i.e. for infinitely slow variations of $b$.

When the wave function is initially prepared in a Landau level before varying $b$, we can intuitively understand why a strictly adiabatic behaviour is not possible: as $b$ varies, the width of the effective harmonic trap changes and thus the wave function is squeezed, which induces a non-zero probability current in the meridional direction $y$, i.e. a non-uniform phase of the wave function. Therefore, the intermediate state cannot correspond to any Landau level, since the latter do not have non-uniform phase factors. But what if the parameter $b$ executes a cycle, i.e. eventually goes back to the initial value $b_1 = b_0$? This situation is the most meaningful regarding the questions of quantum non-adiabaticity and loss of information of the wave function.

Unfortunately, the conventional approach adopted in this section, which consisted in projecting the solution on a basis of time-dependent Landau levels, does not allow us to establish a conclusion, since the local properties and evolution of the wave function are not evident from the perturbative expression \eqref{eq:first-order-adiabatic}, and left incomplete. In contrast, as we will now demonstrate in section \ref{sec:fluid}, the Madelung formulation provides a much clearer perspective regarding this matter, as it will allow us to express quantum non-adiabaticity in terms of non-stationarity of a fluid flow, resulting from out-of-balance effective forces and mechanical energy exchanges.

\section{Fluid mechanics perspective} \label{sec:fluid}

We now introduce the Madelung equations relevant to the present problem. The purpose of this section is to demonstrate how the hydrodynamical perspective provides the exact solution to the time-dependent problem \eqref{eq:Schrodinger_bis}-\eqref{eq:IC_wave_function}, in an alternative way that is straightforwardly interpretable in terms of body forces and local dynamics. The general solution of this problem was originally derived in \cite{lewis1969exact} using quantum invariants, and we propose here an alternative method using the Madelung transform. After reminding that the Madelung equivalent of the Landau levels are shear flows in geostrophic-like balance, we derive an exact solution of the problem. This analysis establishes a direct analogy between quantum non-adiabaticity and the process of adjustment in geophysical flows. Finally, we propose to analyse the result in terms of pseudo-energy of the perturbed flow.

\subsection{Madelung hydrodynamics and Landau levels as geostrophic-like states} \label{part:geostrophic}

Madelung showed in 1927 that the Schrödinger equation can be mapped onto a set of dynamical equations that resemble those of a hypothetical perfect, compressible fluid with a peculiar non-local enthalpy, the Bohm potential. The density and velocity fields in these equations are related to the modulus and the phase gradient of the wave function, solution of the initial Schrödinger equation. Real fluids consist in many particles and their continuous field equations imply the introduction of pressure, which reflects the microscopic interactions between the numerous particles. Conversely, the flow of the Madelung fluid reflects the underlying dynamics of a single quantum particle, which obeys hydrodynamical equations in which the non-local nature of the single particle itself is embedded.

Employing a polar expression of the wave function $\psi (x,y,t) = R \, e^{\ii S/\hbar}$ ($R$ and $S$ are real-valued functions, which can both be positive or negative, and are assumed to be smooth\footnote{Since the evolution of the Madelung fields is enslaved to the Schrödinger equation, then those fields remain well-defined during the evolution, as long as $S$ is sufficiently smooth initially.}), the general equation \eqref{eq:Schrodinger} is mapped onto a set of nonlinear compressible Euler fluid equations. Indeed, defining:
\begin{equation} \label{eq:fields}
    \rho = R^2 \ , \quad \mathbf{u} = \frac{\boldsymbol{\nabla} S - q \mathbf{A}}{m} \ , \quad Q[\rho] = -\frac{\hbar^2}{2 m^2} \frac{\boldsymbol{\nabla}^2 R}{R} = -\left( \frac{\hbar}{2 m} \right)^2 \left[ \boldsymbol{\nabla}^2 \ln \rho + \frac{1}{2} \left( \boldsymbol{\nabla} \ln \rho \right)^2 \right] \ ,
\end{equation}
respectively as the density, velocity field and the Bohm potential associated with the Madelung fluid, the real and imaginary parts of equation \eqref{eq:Schrodinger} yield:
\begin{equation} \label{eq:Euler_equations_mass}
    \partial_t \rho = -\boldsymbol{\nabla} \cdot \left( \rho \mathbf{u} \right) \ ,
\end{equation}
\begin{equation} \label{eq:Euler_equations_HJ}
    -\frac{\partial_t S}{m} = Q + \frac{\mathbf{u}^2}{2} + \frac{q}{m} V \ .
\end{equation}

Equation \eqref{eq:Euler_equations_mass} is the equation of mass conservation and \eqref{eq:Euler_equations_HJ} is the Hamilton-Jacobi equation, which is equivalent to the time-dependent Bernoulli equation of a barotropic flow with enthalpy $Q[\rho]$ \cite{heifetz2015toward}. Taking the gradient of equation \eqref{eq:Euler_equations_HJ} and using the definitions \eqref{eq:Schrodinger} and \eqref{eq:fields} of the electric field $\mathbf{E}$ and the velocity field $\mathbf{u}$, respectively, yield:
\begin{equation} \label{eq:Euler_equations1}
    \partial_t \mathbf{u} = -\boldsymbol{\nabla} \left( Q + \frac{\mathbf{u}^2}{2} \right) + \frac{q}{m} \mathbf{E} \ .
\end{equation}

Furthermore, assuming that the phase $S$ of the wave function is regular\footnote{In this paper we assume that $S$ has no point topological defect, so that the curl of its gradient is zero everywhere. In other situations it may not be so \cite{berry2023time,berry2024kinetically}, in that case the winding points of the wave function generate quanta of non-zero vorticity, i.e. quantised vortices \cite{volovik2003universe}.}, the curl of $\mathbf{u}$ is solely imposed by the magnetic field, as:
\begin{equation} \label{eq:zero_absolute_vorticity}
    \boldsymbol{\nabla} \times \mathbf{u} = -\frac{q}{m} \mathbf{B} = -\boldsymbol{\omega}_c \ ,
\end{equation}
where $\boldsymbol{\omega}_c$ is the cyclotron frequency vector. Besides, recalling the general relation:
\begin{equation}
     \boldsymbol{\nabla} \left(\frac{\mathbf{u}^2}{2}\right) = \left( \mathbf{u} \cdot \boldsymbol{\nabla} \right) \mathbf{u} + \mathbf{u} \times \left( \boldsymbol{\nabla} \times \mathbf{u} \right) \ ,
\end{equation}
we can finally rewrite equation \eqref{eq:Euler_equations1} as an Euler momentum fluid-like equation:
\begin{equation} \label{eq:Euler_equations2}
    \partial_t \mathbf{u} + \left( \mathbf{u} \cdot \boldsymbol{\nabla} \right) \mathbf{u} = -\boldsymbol{\nabla} Q + \frac{q}{m} \left( \mathbf{E} + \mathbf{u} \times \mathbf{B} \right) \ .
\end{equation}

Defining the material derivative of a `fluid parcel' (i.e. an element of the Madelung fluid moving with the flow), $D_t = \partial_t +  \mathbf{u} \cdot \boldsymbol{\nabla}$, the LHS of equation \eqref{eq:Euler_equations2} corresponds to its acceleration while the RHS is the sum of the electromagnetic force (per unit of mass), and the Bohm potential gradient, which acts as an additional force. The Madelung fluid has the following properties:

\begin{itemize}
    \item The fluid is compressible as nothing imposes the divergence of $\mathbf{u}$ to vanish. In fact, in the Landau gauge, the divergence of the magnetic vector potential is zero, so $\boldsymbol{\nabla} \cdot \mathbf{u} = \boldsymbol{\nabla}^2(S/m)$, i.e. the phase can be regarded as a source of compressibility \cite{heifetz2016entropy}. Besides, total mass is conserved as a consequence of normalization of the wave function.
        
    \item The quantum enthalpy or Bohm potential $Q$ follows a barotropic but non-local thermodynamical law, i.e. it depends not only on the fluid's density $\rho$ but also its derivatives \cite{heifetz2015toward}. While the relation between pressure and density in real fluids mirrors their thermodynamical properties, the quantum enthalpy translates wave-particle duality.

    \item The vorticity of the flow (i.e. $\boldsymbol{\nabla} \times \mathbf{u}$) is imposed by the magnetic field, as expressed by equation \eqref{eq:zero_absolute_vorticity}. Under an electromagnetic field, the charged particle is subjected to the Lorentz force, whose magnetic part acts exactly like the Coriolis force for a fluid in a rotating frame\footnote{On a side note, we find it edifying that Madelung hydrodynamics maps the effect of the magnetic field -- which breaks time-reversal symmetry in quantum mechanics -- onto the Coriolis effect, which also breaks time-reversal symmetry in classical fluid dynamics. We believe this has a pedagogical interest, as the fact that rotation breaks time-reversal symmetry might be more comprehensible to a common reader than the similar attribute of magnetic fields in quantum mechanics.}. The cyclotron frequency $\omega_c$ is equivalent to the Coriolis parameter $f$, which is equal to twice the angular frequency of the rotating frame. Viewed from a frame of rest, the total vorticity vector (known as absolute vorticity) is given by $\boldsymbol{\omega}_\text{abs} = \mathbf{f} + \boldsymbol{\nabla} \times \mathbf{u}$. For the Madelung fluid, \eqref{eq:zero_absolute_vorticity} implies that the analogue absolute vorticity is strictly zero, denoted by \cite{heifetz2021zero} as zero absolute vorticity dynamics.   
     
    \item In addition to the magnetic Lorentz force, one can see from equation \eqref{eq:Euler_equations2} that the Madelung fluid undergoes a body force $(q/m) \mathbf{E} = \dot{B} y \mathbf{e}_x$ -- from definitions \eqref{eq:Schrodinger} and \eqref{eq:Landau_gauge} -- which is non-zero whenever the magnetic field varies (by inducing an electric field). As the zero absolute vorticity condition holds as well as when the magnetic field is time-dependent, the rotational part of the velocity $-(q/m) \mathbf{A}(t)$ varies accordingly and can be seen as a result of the work performed by the induced electric field.
\end{itemize}

In the rest of the paper, we will only use dimensionless quantities, as in section \ref{sec:quantum}. The density and dimensionless velocity of the Madelung fluid read as
\begin{equation}
    \rho = |\psi|^2 \quad \text{and} \quad \mathbf{u} =u \mathbf{e}_x + v \mathbf{e}_y = \boldsymbol{\nabla} \arg (\psi) + b y \mathbf{e}_x \ .
\end{equation}

The velocity field is a superposition of a divergent ($\boldsymbol{\nabla} \arg (\psi)$) and a rotational ($b y \mathbf{e}_x$) components. For a spatially independent vertical magnetic field $\mathbf{b} = b \mathbf{e}_z$, the rotational component $u^\text{rot} = b y$ is a plane Couette shear flow, whose vorticity is $\boldsymbol{\omega} =-\partial_y u^\text{rot} \mathbf{e}_z = -\mathbf{b}$, satisfying by construction the zero absolute vorticity condition  \eqref{eq:zero_absolute_vorticity}. It is also convenient to relate the mass flux\footnote{While the velocity is ill-defined where $\psi$ vanishes, the mass flux is always correctly defined. In the Schrödinger picture, its integral yields the expected value of the particle's velocity whereas, in the fluid picture, it yields the net mass transport across the plane.} to the wave function:
\begin{equation}
    \rho \mathbf{u} = \Im \left( \psi^* \boldsymbol{\nabla} \psi \right) + |\psi|^2 b y \mathbf{e}_x \ ,
\end{equation}
where $\Im(z)$ stands for the imaginary part of a complex number $z$ and $z^*$ its conjugate.

Eigenstate solutions of the Schrödinger equation are mapped onto the steady states (stationary solutions, denoted hereafter by overbars) of the continuity \eqref{eq:Euler_equations_mass} and the momentum \eqref{eq:Euler_equations2} equations (i.e. $\partial_t \overline{\rho} = 0$ and $\partial_t \overline{{\mathbf{u}}} = 0$) \cite{heifetz2021zero}. For a constant magnetic field, the eigenstate solutions $\Psi_{n,b} (y)$ defined by \eqref{eq:Fourier_mode}-\eqref{eq:Landau} are translated to
\begin{equation} \label{eq:basic_state}
    \overline{\rho} = \overline{\rho}_{n,b}(y) = | \Psi_{n,b} (y) |^2 \ , \quad \text{and} \quad \overline{\mathbf{u}} = \overline{u}(y) \mathbf{e}_x = b y \mathbf{e}_x \ .
\end{equation}

The continuity equation becomes anelastic (for the terminology, see e.g. section 2.5 of \cite{vallis2017atmospheric}):
\begin{equation} \label{eq:anelastic}
    \boldsymbol{\nabla} \cdot \left( {\overline \rho} \, \overline{\mathbf{u}} \right) = 0 \ ,
\end{equation}
and the Hamilton-Jacobi equation becomes the time-independent Bernoulli equation, with $\omega_{n,b}$ equal to the Bernoulli potential:
\begin{equation} \label{eq:Bernoulli}
    \omega_{n,b} = {\overline Q} + \frac{\overline{\mathbf{u}}^2}{2} \ ,
\end{equation}
where $ \overline{Q}(y) = Q [\overline{\rho}_{n,b}(y)]$, as defined by expression \eqref{eq:fields}. Equation \eqref{eq:Bernoulli} is equivalent to the quantum harmonic oscillator equation \eqref{eq:harmonic_oscillator}, with $\overline{\rho} = \Psi^2$. As opposed to  classical fluids, in the Madelung fluid the Bernoulli potential can only take discrete values. This results from the particular structure of the Bohm potential \eqref{eq:fields}, for which only the specific Hermite-Gauss functions of \eqref{eq:Landau} can set the RHS of \eqref{eq:Bernoulli} to a constant value for the specific shear flow $\overline{u} = b y$. This steady state velocity is exactly equal to the zero absolute vorticity plane Couette flow component. As for all planar shear flows, its nonlinear advection term vanishes ($\left( \mathbf{u} \cdot \boldsymbol{\nabla} \right) \mathbf{u} =0$), leaving the steady state solution in balance between the gradient of the Bohm potential and the magnetic force: $\mathbf{b} \times \overline{\mathbf{u}} = -\boldsymbol{\nabla} \overline{Q}$. As pointed out by \cite{heifetz2021zero}, this is equivalent to the geostrophic balance between the pressure gradient force and the Coriolis force in a rotating shallow water model (Figure \ref{fig:balance}), common to describe geophysical fluid systems  \cite{vallis2017atmospheric}. For the Landau levels \eqref{eq:Landau}, the balance is in the meridional ($y$) direction, while the shear flow is in the zonal ($x$) direction:
\begin{equation} \label{eq:geostrophic_balance}
    b \overline{u} = -\partial_y \overline{Q} \ ,
\end{equation}
as illustrated in Figure \ref{fig:balance}.

\subsection{Non-adiabaticity - deviating from geostrophy by varying \texorpdfstring{$\boldsymbol{b}$}{b}: exact solution} \label{part:adjustment}

Let us now address the time-dependent problem \eqref{eq:Schrodinger_bis}-\eqref{eq:IC_wave_function} through the hydrodynamical formulation. In the scope of this analogy, it corresponds to the following scenario: the initial Madelung flow is in geostrophic balance (Figure \ref{fig:balance}(b)), equation \eqref{eq:geostrophic_balance} ($u(y,t=0) = \overline{u}(y), \, \rho(y,t=0) = \overline{\rho}(y)$), i.e. such that $b_0 \overline{u} = -\partial_y Q[\overline{\rho}]$ (with $\overline{u} = b_0 y$ and $\overline{\rho} = |\Psi_{n_0 , b_0} (y)|^2 = \frac{1}{2^{n_0} n_0 !} \sqrt{ \frac{b_0}{\pi} } \: \ee^{-b_0 y^2} H_{n_0}^2 \left( \sqrt{b_0} y \right)$), then imbalance is induced by changing the value of $b$ in time, which breaks the equilibrium between the Lorentz force and the gradient of the Bohm potential. For the one-dimensional ($y$) time-dependent Schrödinger equation \eqref{eq:Schrodinger_bis}, the density and velocity fields are independent of $x$ at any time, thus the corresponding two components of the Madelung momentum equation \eqref{eq:Euler_equations2} and continuity equation \eqref{eq:Euler_equations_mass} read as:
\begin{equation} \label{eq:Euler_1D}
    \begin{split}
        \partial_t u &= \left( b(t) - \partial_y u \right) v + \dot{b}(t) y \ , \\[6pt]
        \partial_t v &= -v \partial_y v - \partial_y Q - b(t) u \ , \\[6pt]
        \partial_t \rho &= -\partial_y (\rho v) \ .
    \end{split}
\end{equation}

The zero absolute vorticity constraint yields immediately that the zonal momentum  $u$ is equal to $b(t) y$ at all times. This implies an instantaneous\footnote{Classical systems do not usually respond instantaneously to a change of external parameters.} balance between the zonal component of the Lorentz force and the zonal momentum advection by the meridional component of the flow (so that $\left( b(t) - \partial_y u \right) v=0$). Thus, the zonal component of the flow is changing only through the induced electric field $\dot{b} y \mathbf{e}_x$. The second and third of equations \eqref{eq:Euler_1D} couple the density variations with the meridional flow $v$. Using the material derivative $D_t = \partial_t + v \partial_y$, they can be equivalently written as:
\begin{equation}
    \begin{split} \label{eq:1D_dynamics}
       &D_t v = -\partial_y Q - b(t) u = -\partial_y Q - b^2(t) y \ , \\[6pt]
       &D_t \ln{\rho} = -\partial_y v \ .    
    \end{split}
\end{equation}

\begin{figure}[t]
\begin{center}
\includegraphics[scale=0.55]{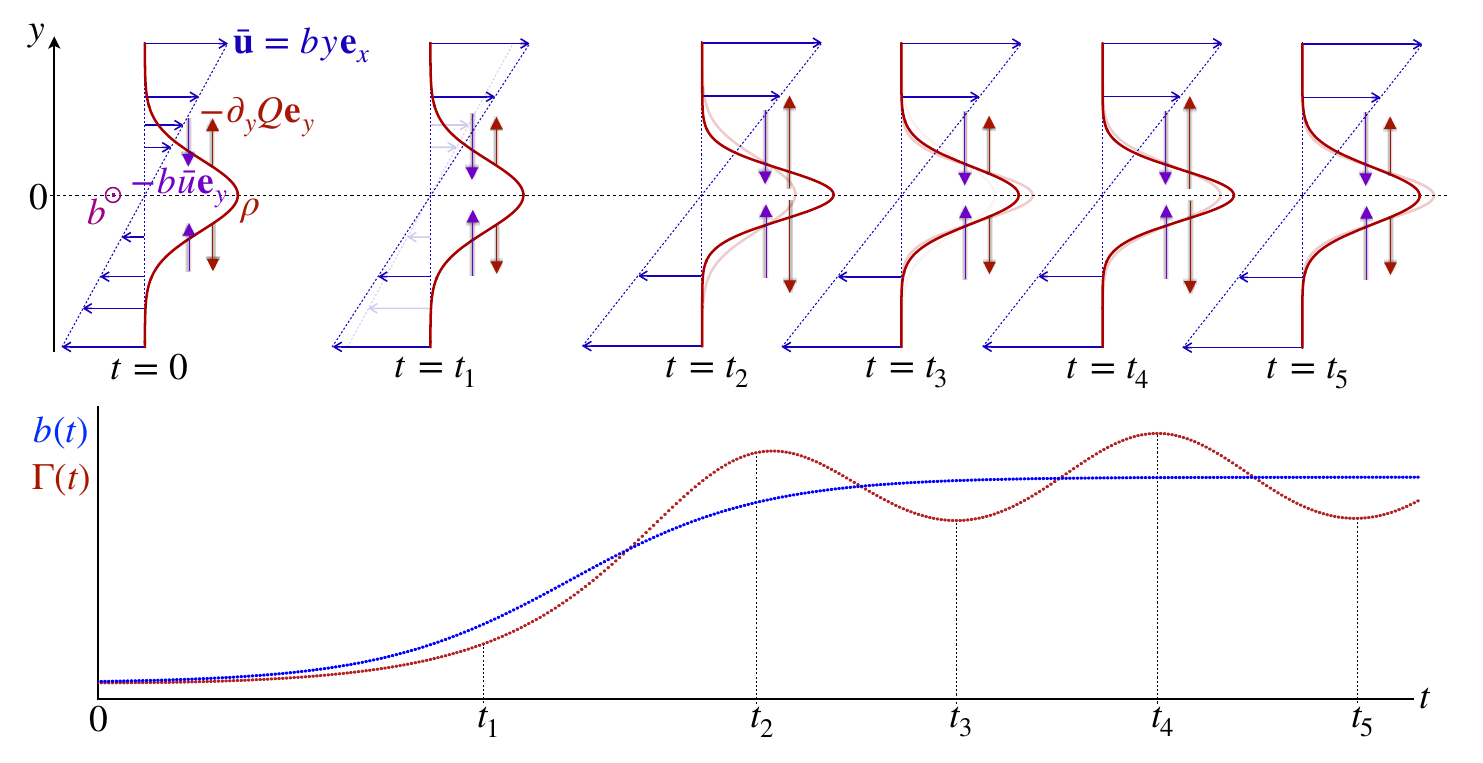}
\end{center}
\caption{\label{fig:mechanisms} Illustration of the restoring mechanisms at play during the transverse adjustment process. At $t=0$, the Lorentz force and the gradient of potential $Q$ are balanced. As $b$ varies (increases in this example), the amplitude of the zonal Couette flow, and thus the meridional Lorentz force, immediately change ($t=t_1$). This induces a meridional flow with a delay, which redistributes density ($t=t_2$) toward the centre, which in turns changes the gradient of $Q$ to act against the inward convergence in the meridional direction. However, the dynamics does not converge to a new geostrophically-balanced state \eqref{eq:geostrophic_balance}, due to inertia ($t=t_3$), even when $b$ converges to a constant value  ($t \geq t_4$). In the absence of dissipation, everlasting oscillations of the density and meridional velocity - the signature of the non-adiabatic process -  whose frequency is given by twice of the final cyclotron frequency, are accompanying the newly established geostrophic balance.}
\end{figure}

Initially, we have $v = 0$ since the Couette flow is in geostrophic balance ($-\partial_y Q [\overline{\rho}] - b_0 \overline{u} = 0$). As $b$ begins to vary (say to increase), a non-zero meridional velocity arises and converges the density toward the centre, which in turn alters $-\partial_y Q [\rho]$, as demonstrated in Figure \ref{fig:mechanisms}. Since the initial deviation from geostrophic balance is due to the increase in the magnetic Lorentz force, whose meridional component is proportional to $y$, we can predict that the initial meridional acceleration of a fluid parcel at position $y$ is also proportional to $y$. Moreover, since $D_t v = \partial_t v + \partial_y (v^2/2)$, we expect the generated flow to preserve the simple structure:
\begin{equation} \label{eq:v}
    v = \beta (t) y \ ,
\end{equation}
which is in agreement with the ansatz proposed by \cite{sutherland1998exact} for coherent states evolution. The continuous response of `fluid parcels' thus exhibits the following properties:
\begin{equation} \label{eq:fluid_parcel_conservation}
    \begin{split}
        D_t y = v = \beta(t) y \quad &\Rightarrow \quad y(t) = y(0) \exp \left\{ \int_{0}^t \dd t' \, \beta(t') \right\} \ , \\[6pt]
        D_t {\rho} = -\beta(t)  {\rho}\quad &\Rightarrow \quad \rho(y(t),t) = \rho(y(0),0) \exp \left\{ -\int_{0}^t \dd t' \, \beta(t') \right\} \ .
    \end{split}
\end{equation}

Therefore, the density field is self-similar in time and reads as:
\begin{equation} \label{eq:adjusted_density}
    \begin{split}
        \rho(y,t) &= \rho \left( y \exp \left\{ -\int_{0}^t \dd t' \, \beta(t') \right\} , 0 \right) \: \ee^{- \int_{0}^t \dd t' \beta(t')} \\[6pt]
        &= \frac{1}{2^{n_0} n_0 !} \sqrt{ \frac{b_0}{\pi} \ee^{- 2\int_{0}^t \dd t' \beta(t')}} \: \exp \left\{-y^2 \: b_0 \ee^{-2 \int_{0}^t \dd t' \beta(t')} \right\} H_{n_0}^2 \left( y \: \sqrt{b_0 \ee^{- 2\int_{0}^t \dd t' \beta(t')}} \right) \ .
    \end{split}
\end{equation}

Remarkably, the properties \eqref{eq:fluid_parcel_conservation} imply that the product $(y\rho)$ is conserved with a fluid parcel's motion ($D_t (y \rho) = 0$) as $b(t)$ varies\footnote{In particular, if the initial state has lines of zero density, i.e. where $\rho=0$ (this is the case if $n_0 \geq 1$), these are preserved and move with the fluid parcels.}. Plugging \eqref{eq:v} in the first of equations \eqref{eq:1D_dynamics}, we obtain:
\begin{equation} \label{eq:Bohm_beth}
    \partial_y Q + \left( \dot{\beta} + \beta^2 + b^2 \right) y = 0 \quad \Rightarrow \quad Q + \left( \dot{\beta} + \beta^2 + b^2 \right) \frac{y^2}{2} = \alpha (t) \ .
\end{equation}

In other words, at any time $t$, $R(y,t)$ is an eigenfunction of the quantum harmonic oscillator with the time-dependent trapping frequency:
\begin{equation} \label{eq:beth}
    \Gamma (t) = \sqrt{ \dot{\beta} + \beta^2 + b^2 } \ ,
\end{equation}
and eigenvalue $\alpha (t)$. Since the initial conditions are $R(y,0) = \Psi_{n_0 , b_0} (y)$ and $\Gamma (0) = b_0$, we thus have:
\begin{equation} \label{eq:density}
    R(y,t) = \Psi_{n_0 , \Gamma (t)} (y) \quad \text{and} \quad \alpha (t) = \Gamma(t) \left( n_0 + \frac{1}{2} \right) \ .
\end{equation}

Expression \eqref{eq:density}, which is exact\footnote{Keep in mind that this does not solve completely equation \eqref{eq:Schrodinger_bis}, as both $\beta (t)$ and the phase of the wave function are yet to be found.}, would not have been obtained through the somewhat naive decomposition \eqref{eq:Berry_projections} of the wave function, yet both formulations are insightful as they exhibit an exact equivalence between the non-adiabatic response of the quantum system and the ageostrophic dynamics of the Madelung fluid. Indeed, a flow that is constantly adjusted to the instantaneous value of $b(t)$ would read $\Gamma (t) = b(t)$, which also corresponds to the adiabatic evolution of the wave function. Nevertheless, even if the variation of the magnetic field is quasi-adiabatic, an initial plane-Couette stationary state cannot remain exactly stationary, for the following simple reason: as $b$ increases or decreases, the density field of the corresponding Landau level is compressed or stretched in the meridional direction as the intensity of the zonal flow -- and thus of the meridional Lorentz force -- adjusts to the new value of $b$ (adiabatic expansion). This can only happen with a non-zero velocity field in $y$, which means that the flow is not in geostrophic balance during the transition. Besides, in the expression \eqref{eq:beth} of $\Gamma (t)$, the term $b^2(t)$ comes from the adjustment of the Lorentz force\footnote{Note that the meridional Lorentz force is modified according to both the variation of $b(t)$ and the corresponding instantaneous variation of $u = b(t) y$.}, whereas the term $\dot{\beta} + \beta^2$ comes from the subsequent variations of $v$ (inertia). In other words, the difference between $\Gamma (t)$ and $b (t)$ -- which characterises the non-adiabatic evolution of the quantum system -- originates from the non-zero meridional velocity, which also embodies the ageostrophic dynamics of the Madelung fluid.

Comparing the solution $\rho(y,t) = |\Psi_{n_0 , \Gamma(t)}(y)|^2$ with expression \eqref{eq:adjusted_density} allows us to identify:
\begin{equation} \label{eq:useful_relation}
    \Gamma^2 (t) = \dot{\beta} + \beta^2 + b^2 = b_0^2 \exp \left\{ -4\int_{0}^t \dd t' \, \beta(t') \right\} \ ,
\end{equation}
therefore we obtain a second-order differential equation\footnote{Equation \eqref{eq:beta} can be regarded as a remote variant of the Duffing equation, describing a nonlinear oscillator whose linear component has a time-dependent coefficient (parametric oscillator) and the forcing has the general form of $-\dot{b^2}$. A major difference is that the term $6 \dot{\beta} \beta = 3\dot{\beta^2}$ cannot be considered as a damping term.} for $\beta(t)$:
\begin{equation} \label{eq:beta}
    \ddot{\beta} + 4 b^2 \beta + 6 \dot{\beta} \beta + 4 \beta^3 + 2 \dot{b} b = 0 \ .
\end{equation}

\begin{figure}[t]
\begin{center}
\includegraphics[scale=0.5]{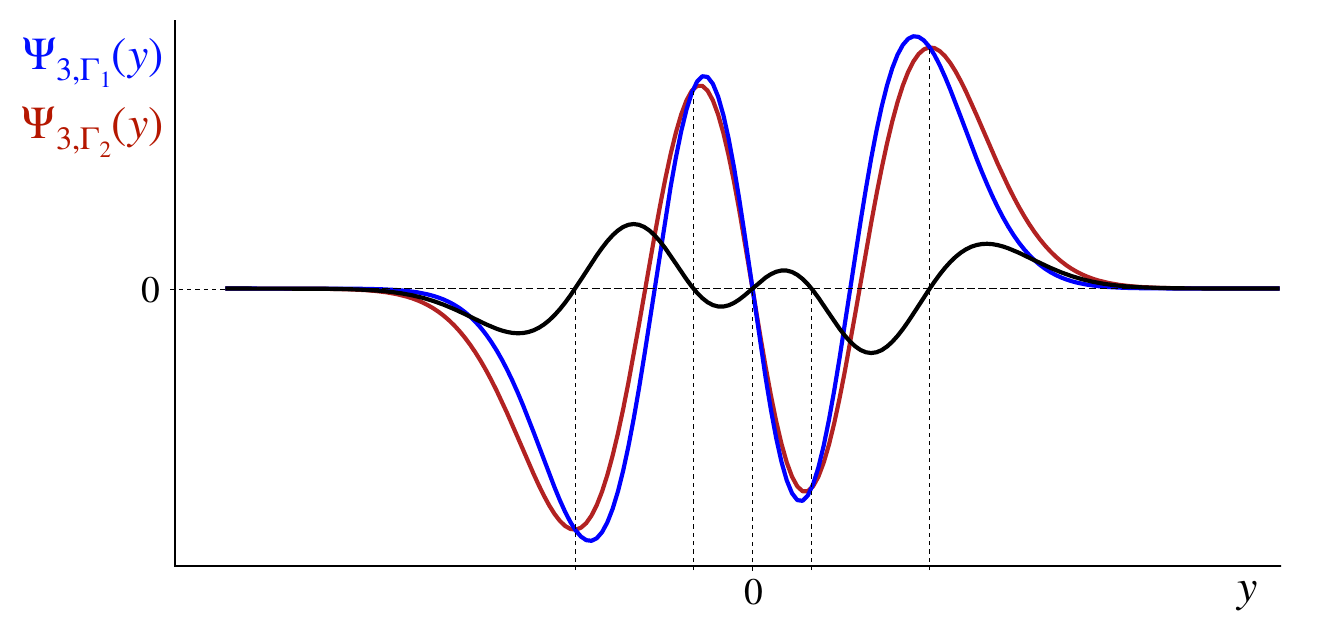}
\end{center}
\caption{\label{fig:perturbation} Stretching of the amplitude $R(y,t)$ as $\Gamma$ changes from a value $\Gamma_1$ (dark blue curve) to a slightly smaller value $\Gamma_2$ (dark red curve), in the case $n = 3$. The black curve represents the difference between the two amplitudes, which has $n + 2 = 5$ zeros owing to the crossings between the initial amplitude function and the stretched one. This result is consistent with the perturbative approach -- since the obtained expression for the perturbed wave function involves $\Psi_{n+2}$ at lowest order (see section \ref{sec:quantum} and appendix \ref{apx:match}) -- but the latter does not reveal that the amplitude is at all times the same rescaled Hermite-Gauss function. Conversely, the hydrodynamic formulation presented in section \ref{sec:fluid} naturally demonstrates this self-similar character of the amplitude at all times.}
\end{figure}

We can observe that, while the velocity field $\mathbf{u} = b(t) y \mathbf{e}_x + \beta(t) y \mathbf{e}_y$ is independent of the level number $n_0$, $R(y,t)$ remains the $n_0^\text{th}$ Hermite-Gauss function (Figure \ref{fig:perturbation}), so the density field has $n_0$ zeros in the meridional direction and vanishes away from the centre over a typical distance $y \sim 1/\sqrt{\Gamma}$, which characterises the response of the Madelung fluid -- initially in a state of stationary zonal flow -- to the variation of $b(t)$. Since $\beta(0) = 0$ and $\dot{\beta}(0) = 0$ for a regular function $b(t)$, the driving term $2 \dot{b} b$ is the one that produces non-zero meridional velocity. Let us note that the solution of equation \eqref{eq:beta} -- and thus the function $\Gamma (t)$ as well -- is completely determined by $b(t)$ and is independent of the level index $n_0$. Finally, the expression of the full wave function\footnote{The phase $s$ of the wave function is determined through the definition $\boldsymbol{\nabla} s = v \mathbf{e}_y$ and the Bernoulli equation $-\partial_t s = Q + \mathbf{u}^2 / 2$.}, solution of \eqref{eq:Schrodinger_bis}-\eqref{eq:IC_wave_function}, is
\begin{equation} \label{eq:wave_function_solution}
    \psi (y,t) = \exp \left\{ \ii \beta(t) \frac{y^2}{2} - \ii \left( n_0 + \frac{1}{2} \right) \int_{0}^t \dd t' \, \Gamma (t') \right\} \Psi_{n_0 , \Gamma(t)} (y) \ ,
\end{equation}
where $\beta$ and $\Gamma$ are respectively obtained by solving equation \eqref{eq:beta} and using relation \eqref{eq:beth}. It is worth noting the connection between our exact solution \eqref{eq:wave_function_solution} and the standard quantum invariant method. In the Lewis-Riesenfeld framework \cite{lewis1969exact}, the evolution of the time-dependent harmonic oscillator relies on an auxiliary scaling parameter $\varrho(t)$ that satisfies the nonlinear Ermakov equation, i.e. $\ddot{\varrho} + b(t)^2 \varrho = b_0^2 \varrho^{-3}$, with the initial conditions $\varrho(0)=1$ and $\dot{\varrho}(0)=0$. In our Madelung formulation, the geometric constraint of zero absolute vorticity dictates the exact physical fluid flow, naturally yielding the time-dependent trapping frequency $\Gamma (t)$. The typical width of the density field scales as $\Gamma^{-1/2}$, thus the variable $\Gamma$ plays the exact physical role of\footnote{Actually, we have the correspondences $\Gamma(t) = b_0 \varrho(t)^{-2}$ and $\beta(t) = \dot{\varrho}/\varrho$, which are compatible with expression \eqref{eq:wave_function_solution} and equation \eqref{eq:useful_relation}.} $\varrho^{-2}$, thereby bypassing the need to postulate an auxiliary function and instead deriving relations \eqref{eq:useful_relation} and the parametric oscillator equation \eqref{eq:beta} directly from the conservation of momentum and mass in the perturbed Couette flow.

Let us now examine the permanent regime (constant $b$), say after varying $b(t)$ to a new value $b_1 \neq b_0$. Equation \eqref{eq:beta} then becomes a nonlinear equation of free oscillations:
\begin{equation} \label{eq:beta_free}
    \underbrace{\ddot{\beta} + 4 b_{1}^2 \beta}_\text{linear oscillations} + \underbrace{6 \dot{\beta} \beta + 4 \beta^3}_\text{nonlinear terms} = 0 \ .
\end{equation}

The linear part in equation \eqref{eq:beta_free} characterises an oscillator of frequency $2 b_1$, which is in agreement with the conclusions of section \ref{sec:quantum}. The second term, which indicates the frequency of the persistent oscillations, arises from the Lorentz force -- proportional to $b^2$ -- and the properties of $Q$. While equation \eqref{eq:beta} for the transiting regime cannot be solved analytically in general, there is an exact solution of equation \eqref{eq:beta_free} for the permanent regime's dynamics, and one can check (see appendix \ref{apx:beta}) that it is given by\footnote{This solution is consistent with expression (13) found in \cite{sutherland1998exact} for the coherent sloshing mode. As discussed in the introduction, the one-dimensional quantum problem \eqref{eq:Schrodinger_bis} was analysed by \cite{sutherland1998exact} in the context of a more general $N$-body system with additional pair interactions and a solution was proposed for the coherent states. While the solution of \cite{sutherland1998exact} is derived after assuming an ansatz for the wave function, here the solution results directly from the Madelung formulation.}:
\begin{equation} \label{eq:exact_solution}
    \begin{split}
        \beta (t) &= \frac{- b_1 \epsilon \sin \left( 2 b_1 t + \varphi \right)}{1 + \epsilon \cos \left( 2 b_1 t + \varphi \right)} \ , \\[6pt]
        \text{thus} \quad \Gamma (t) &= \frac{b_1 \sqrt{1 - \epsilon^2}}{1 + \epsilon \cos \left( 2 b_1 t + \varphi \right)} \ ,
    \end{split}
\end{equation}
where the constants $\epsilon, \varphi$ are given by the value of $\beta$ and $\dot{\beta}$ at some time to solve the Cauchy problem. In other words, expressions \eqref{eq:exact_solution} are the exact solution after varying $b(t)$ to a constant value $b_1$, and $\epsilon,\varphi$ depend on the past evolution of $b(t)$. The smallness of the coefficient $\epsilon$ -- which is the Madelung analog of the Rossby number \cite{vallis2017atmospheric, heifetz2021zero} -- quantifies the adiabatic character of the quantum system. For instance, if $b(t)$ is a step function that is equal to $b_0$ for $t \leq 0$ and $b_1$ for $t > 0$\footnote{Let us note that this situation is the equivalent of having a constant value $b(t) = b_1$ at all time and start with an eigenfunction $\Psi_{n_0 , b_0}$ with some $b_0 \neq b_1$, which is a non-stationary state.}, then the velocity field for $t > 0$ is given by expressions \eqref{eq:exact_solution} with $\varphi = 0$ and $\epsilon = \left( b_{1}^2 - b_0^2 \right) / \left( b_{1}^2 + b_0^2 \right)$\footnote{Note that $\epsilon$ is indeed a small parameter for a small variation of $b$, and its sign is in agreement with the expected sign of the meridional velocity field right after $t=0$: for instance, if $b_1 > b_0$, $v$ is negative for $y>0$ and positive for $y<0$, which is compatible with the compression of the density toward the centre as the Lorentz force increases.}, which are compatible with $\beta (0) = 0$ and $\Gamma (0)^2 = b_0^2 = b_{1}^2 + \dot{\beta}(0)$. Expressions \eqref{eq:exact_solution} reveal that the Madelung flow adopts a periodic behaviour in the permanent regime, and that the frequency of the oscillations\footnote{Note that expanding the denominator in the expression of $\beta$ provides an explicit decomposition in a Fourier series.} is given by $2 b_1$, as alternatively concluded in section \ref{sec:quantum}. These periodic global oscillations of the Madelung fluid constitute a sloshing mode, which is the result of the restoring mechanisms of the Madelung fluid, owing to the Lorentz force and the gradient of Bohm potential (Figure \ref{fig:mechanisms}). The slower is the variation of $b(t)$, the smaller is the amplitude of these oscillations, however expressions \eqref{eq:wave_function_solution} and \eqref{eq:exact_solution} are exact in the permanent regime, no matter the rate of change of $b(t)$. Eventually, the amplitude of the oscillations of $\beta$ is entirely determined by the memory of the transition through equation \eqref{eq:beta}. This solution in the permanent regime naturally emerges as a consequence of the hydrodynamical interpretation of the adjustment problem, characterising quantum non-adiabaticity and providing an interpretation for the frequency of the reminiscent oscillations as a manifestation of interlevel transitions. In appendix \ref{apx:match}, we show that the exact expression \eqref{eq:wave_function_solution} matches the approximate result \eqref{eq:first-order-adiabatic} in the adiabatic limit.

\subsection{Sloshing energetics and pseudo-energy} \label{part:energy}

In the previous analysis, we showed that any change of $b(t)$ to a new constant value leads to a new Madelung flow that is the superposition of the adjusted stationary flow and a persistent oscillation in the meridional direction. In terms of energetics, during the adjustment process, there are two sources of available energy: the energy injected in the flow through the variation of $b(t)$\footnote{More specifically, energy is injected to the flow via the work of the induced field that appears in the zonal momentum equation \eqref{eq:Euler_1D}.} with time, and the initial background flow energy, which is the sum of Bohm's potential energy and zonal kinetic energy. To understand the mechanisms of energy partitioning between the oscillation and the adjusted mean flow, let us identify and compute these different contributions. Generally speaking, the energy of the Madelung flow can be expressed as\footnote{Let us remind that we consider flows that do not depend on the zonal coordinate $x$.}:
\begin{equation} \label{eq:energy_def}
    E = \int \left( Q + \frac{\mathbf{u}^2}{2} \right) \rho \, \dd y = \int \psi^* \mathcal{H} \psi \, \dd y \ ,
\end{equation}
where the operator $\mathcal{H}$ is the one expressed in equation \eqref{eq:Schrodinger_bis}. Therefore, the total amount of energy absorbed by the system (provided by the induced electric field, i.e. the external variation of $b$) per unit of time is equal to:
\begin{equation} \label{eq:power_expr}
    \frac{\dd E}{\dd t} = \int \psi^* \partial_t \mathcal{H} \psi \, \dd y = b(t) \dot{b}(t) \int \rho(y,t) y^2 \, \dd y \ .
\end{equation}

We can make a few comments about expression \eqref{eq:power_expr}. In the Madelung fluid perspective, it corresponds to the work of the induced electric field $\mathbf{E} = \dot{b} y \mathbf{e}_x$ on all the fluid parcels of mass $\rho(y) \dd y$ and zonal velocity $u = b y$, per unit of time. If $\rho(y)$ were not varying in time, then all the work would be converted into zonal kinetic energy, and if $b(t)$ were to execute a cycle -- i.e. come to its initial value -- then the total work would be zero, as the integral of \eqref{eq:power_expr} would vanish. However, we can see that the work injected not only depends on the field strength ($b$ and $\dot{b}$), but also on the mass distribution during the time work is performed -- i.e. the function $\rho(y,t)$ -- which is characteristic of a parametric oscillator. Owing to this property, we can already expect situations of energetic hysteresis, in which the net injection of work is non-zero even if $b(t)$ eventually comes back to its initial value. Starting in the Landau state $n_0$ for the initial value $b(t=0) = b_0$, we can show (see appendix \ref{apx:proofs}) that the rate of energy injection is:
\begin{equation} \label{eq:injected_energy}
    \frac{\dd E}{\dd t} = \left( n_0 + \frac{1}{2} \right) \frac{b(t) \dot{b}(t)}{\Gamma (t)} \ .
\end{equation}

\begin{figure}[t]
\begin{center}
\includegraphics[scale=0.8]{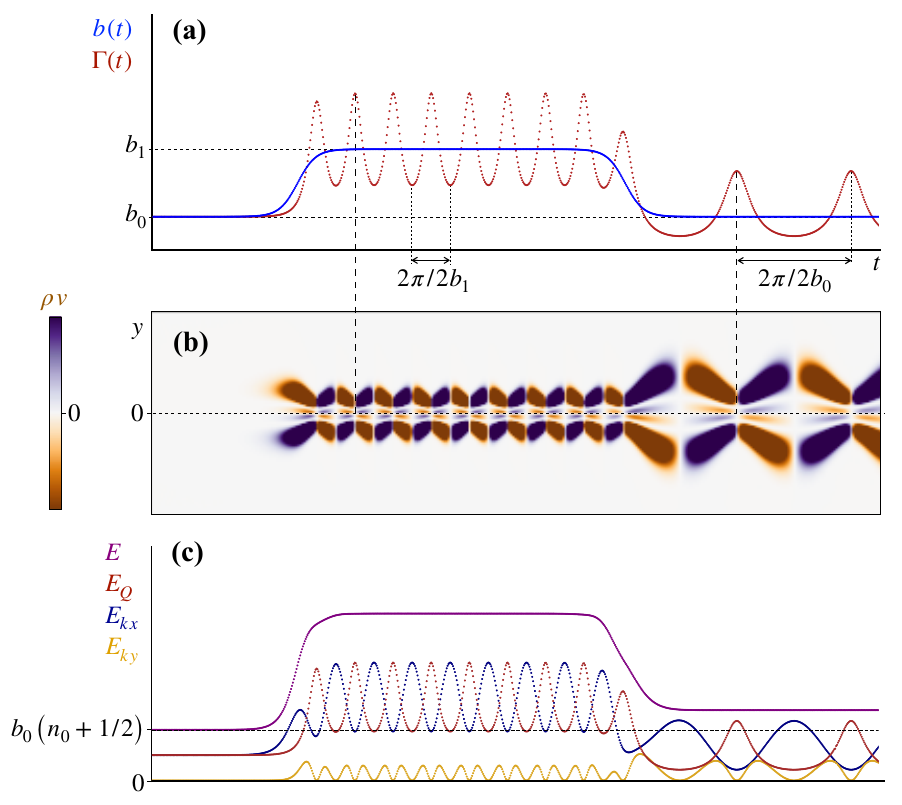}
\end{center}
\caption{\label{fig:b_VS_beth} Response of the Madelung fluid to a variation of $b$. \textbf{(a)} Evolution of $\Gamma (t)$ in response of $b(t)$. In the transition phases, we can observe the delay of the response, which then catches up with $b(t)$, thus generating an oscillatory behaviour, whose frequency is given by $2b$ in the phase of constant $b$. \textbf{(b)} Plot of the meridional mass flux $\rho v$ in time. We picked the Landau level $n_0 = 2$. The global zeros coincide with the moments when the density is maximally and minimally squeezed (vertical dashed lines between plots \textbf{a} and \textbf{b}). In the permanent regime, the oscillations persist even if $b(t)$ is eventually set back to its initial value, which reveals a hysteresis phenomenon. \textbf{(c)} Plot of the total energy $E$ in function of time, and the partition between kinetic energy and Bohm's potential energy. The hysteresis appears clearly as the total energy does not return to its original value, leaving a positive difference which is the energy of the persistent sloshing oscillations. Numerical methods and the values of the parameters are addressed in appendix \ref{apx:numerics}.}
\end{figure}

At initial time $t=0$, the total energy of the flow is equal to $(n_0 + 1/2) b_0$ and it is equally partitioned between zonal kinetic energy and Bohm's potential energy \cite{heifetz2021zero}. This is not true anymore for $t>0$. Indeed, because of the properties of the function $\Psi_{n_0 , \Gamma (t)}$, we have\footnote{Expressions \eqref{eq:energy_partition} reveal that equipartition between zonal kinetic and potential energy is possible only if $\Gamma (t) = b(t)$, which would correspond to the exact adiabaticity of the quantum system.} (see appendix \ref{apx:proofs}):
\begin{equation} \label{eq:energy_partition}
    \begin{split}
        E_{kx} &= \left( n_0 + \frac{1}{2} \right) \frac{b(t)^2}{2 \Gamma (t)} \quad \text{(zonal kinetic energy)} \ , \\[6pt]
        E_{ky} &= \left( n_0 + \frac{1}{2} \right) \frac{\beta(t)^2}{2 \Gamma (t)} \quad \text{(meridional kinetic energy)} \ , \\[6pt]
        E_{Q} &= \left( n_0 + \frac{1}{2} \right) \frac{\Gamma (t)}{2} \quad \text{(potential energy)} \ .
    \end{split}
\end{equation}

At time $t=0$, the flow is initially at rest, thus $E_{ky} = 0$ and the total energy, i.e. $E = E_{kx} + E_{ky} + E_Q$, is equal to $(n_0 + 1/2) b_0$. For $t \geq 0$, as $b$ varies, we have
\begin{equation} \label{eq:positive_energy}
    E - \left( n_0 + \frac{1}{2} \right) b(t) = \frac{n_0 + 1/2}{2 \Gamma} \left( \beta^2 + (\Gamma - b)^2 \right) \geq 0 \ ,
\end{equation}
which means that the energy of the flow is systematically higher than the expectation energy of the adjusted mean flow. In other words, there is always a positive amount of energy which corresponds to the generation of persistent oscillations (Figure \ref{fig:b_VS_beth}). Energy is injected to or extracted from the flow by the external action of the induced zonal field $\dot{b}(t) y \mathbf{e}_x$, and expression \eqref{eq:injected_energy} shows that the sign of $\dot{b}$ determines whether the energy is injected or extracted. If $b(t)$ executes a cycle such that its initial and final values are equal, the additional energy in \eqref{eq:positive_energy} does not vanish in general, which is symptomatic of a non-adiabatic evolution of the quantum wave function. Rather than a perfect adjustment of the flow to a new geostrophic-like state, we thus observe an energetic hysteresis owing to the systematic generation of these oscillations, which may not be reversed simply by returning $b$ to its initial value. This hysteresis phenomenon thus characterises the non-adiabaticity of the quantum system in terms of global waves persisting in the Madelung fluid.

\begin{figure}[t]
\begin{center}
\includegraphics[scale=0.8]{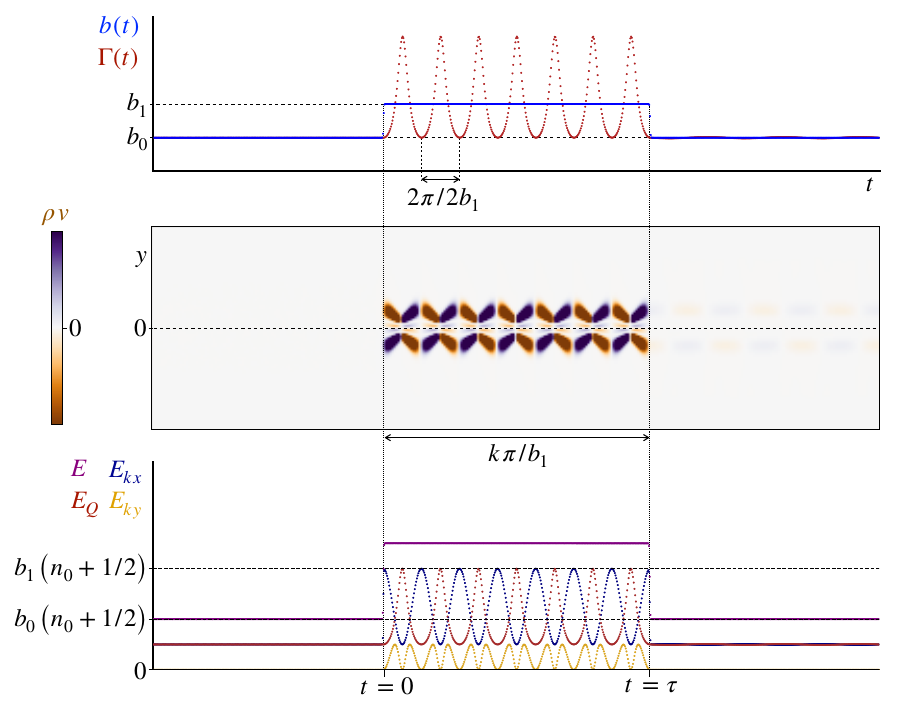}
\end{center}
\caption{\label{fig:reversibility} Similar situation as the one depicted in Figure \ref{fig:b_VS_beth}, only here $b$ varies nearly step-wise and returns to its initial value $b_0$ at the exact moment when $\Gamma = b_0$. In this marginal situation, sloshing oscillations are nearly suppressed and $\Delta E$ vanishes. Numerical methods and the values of the parameters are addressed in appendix \ref{apx:numerics}.}
\end{figure}

The final state is a geostrophic one if and only if expression \eqref{eq:positive_energy} vanishes, i.e. $\beta = 0$ (no transverse motion) and $\Gamma = b$, which can be achieved only in marginal cases depending on how the variations of $b(t)$ are synchronised with the oscillations of the flow. Let us consider a simple example in which $b(t)$ varies stepwise, going from $b_0$ to $b_1 \neq b_0$ at time $t=0$ and back to $b_0$ at time $t=\tau$. For the flow to come back to its initial state, i.e. for the oscillations initiated at $t=0$ to be suppressed, the total energy balance must be zero, i.e. the integral of expression \eqref{eq:power_expr}. Since energy is injected or removed only at times $t=0$ and $\tau$, this integral can be expressed as:
\begin{equation}
    \Delta E = \left( n_0 + \frac{1}{2} \right) \left( \frac{b_1^2 - b_0^2}{2} \right) \left( \frac{1}{\Gamma (0)} - \frac{1}{\Gamma (\tau)} \right) \ ,
\end{equation}
which is zero if and only if $\tau$ is chosen such that $\Gamma (\tau) = \Gamma (0)$ (Figure \ref{fig:reversibility}). This is consistent with the general form \eqref{eq:exact_solution} of the solution on intervals of constant $b$: since $\beta$ and $\Gamma$ are continuous, $b$ must be returned to the initial value $b_0$ when $\Gamma(\tau) = b_0$, which coincides with the cancellation of meridional velocity\footnote{During the time interval $0 < t < \tau$, the function $\Gamma$ oscillates between the values $b_0$ and $b_1^2 / b_0$ with frequency $2 b_1$, as can be noticed from expression \eqref{eq:exact_solution}, so $b_0$ is an extreme value of $\Gamma$.} and times $\tau = k \pi / b_1$ with integer $k$. This behaviour can be related to that of classical parametric oscillators, for which some property of the system can grow or decay depending on how the variations of a parameter are synchronised with the oscillations of the system itself. It is important to note that the persistent sloshing oscillations thus vanish only in marginal situations, when $b$ is eventually returned to its original value in a way that equilibrates the energy exchanges between the flow and the external excitation.

Alternatively, we can compute the pseudo-energy of the oscillations under the linear approximation. We assume that the permanent regime is established ($\dot{b} = 0$ and $b = \overline{b}$) and the Madelung flow is the superposition of the corresponding geostrophic-like state and some perturbation\footnote{As stressed in \cite{heifetz2023broglie}, any linear combination of wave functions that solve the Schrödinger equation is still a solution of it, however the superposition of their respective Madelung flows is not a solution of the hydrodynamical equations, in general.}, i.e. $v = v'$ and $\rho = \overline{\rho} + \rho'$ with $\overline{\rho}(y) = |\Psi_{n_0 , \overline{b}} (y)|^2$. We can thus write the total energy as:
\begin{equation}
    E = \int \left( \overline{Q} + Q' + \overline{b}^2 \frac{y^2}{2} + \frac{v'^2}{2} \right) \left( \overline{\rho} + \rho' \right) \, \dd y \ .
\end{equation}

By definition, $\overline{Q} + \overline{b}^2 y^2 / 2$ is equal to $(n_0 + 1/2) \overline{b}$, which is the energy of the unperturbed geostrophic flow. Using expression \eqref{eq:fields} and the Taylor expansion $\ln(\overline{\rho} + \rho') = \ln \overline{\rho} + \left( \rho'/\overline{\rho} \right) - \left( \rho' / \overline{\rho} \right)^2 / 2 + \mathcal{O} \left( \left( \rho' / \overline{\rho} \right)^3 \right)$, Bohm's potential can be written as:
\begin{equation} \label{eq:Q-expansion}
    Q = \overline{Q} -\frac{1}{4} \left[ \partial_{yy} \frac{\rho'}{\overline{\rho}} + \left( \frac{\dd}{\dd y} \ln \overline{\rho} \right) \partial_y \frac{\rho'}{\overline{\rho}} - \frac{1}{2} \partial_{yy} \left( \frac{\rho'}{\overline{\rho}} \right)^2 - \frac{1}{2} \left( \frac{\dd}{\dd y} \ln \overline{\rho} \right) \partial_y \left( \frac{\rho'}{\overline{\rho}} \right)^2 + \frac{1}{2} \left( \partial_y \frac{\rho'}{\overline{\rho}} \right)^2 \right] \ ,
\end{equation}
up to order square in the perturbation fields, for the purpose of computing the pseudo-energy. Bearing in mind that conservation of mass implies $\int \rho' \, \dd y = 0$\footnote{Indeed, $\int \rho \, \dd y = \int \overline{\rho} \, \dd y = 1$.}, and keeping terms only up to quadratic order in the perturbations, we can show the following expression of the pseudo-energy:
\begin{equation} \label{eq:pseudo-energy}
    E - \left( n_0 + \frac{1}{2} \right) \overline{b} = \frac{1}{2} \int \left[ v'^2 + \left( \frac{1}{2} \partial_y \frac{\rho'}{\overline{\rho}} \right)^2 \right] \overline{\rho} \, \dd y \ .
\end{equation}

Thus, up to quadratic order, the pseudo-energy is a constant of motion (in the permanent regime), which is positive definite in the perturbations, consistently with \eqref{eq:positive_energy}\footnote{However, expression \eqref{eq:pseudo-energy} is valid for general small perturbations of the permanent stationary flow, not only the situation in which $b(t)$ varies to a different permanent regime, which yields expression \eqref{eq:positive_energy}.}. Hence, the Madelung flow is stable to perturbations, a fact that could be anticipated from the start owing to the Hermiticity of the quantum system. Expression \eqref{eq:pseudo-energy} is characteristic of the Madelung fluid, as the potential contribution (i.e. the second term in the integral) depends on the derivative of the density perturbation, and is thus non-local. Expression \eqref{eq:pseudo-energy} is the counterpart of the expression (22) derived in \cite{heifetz2020madelung}, and it represents the energy of a small perturbation of a stationary plane-Couette flow, in particular of the persistent sloshing oscillation accompanying the adjustment process, whose energy is alternatively exactly given by \eqref{eq:positive_energy}. Nevertheless, if the background density $\overline{\rho} (y)$ has zeros (i.e. for $n_0 \geq 1$), the well-definiteness of the Taylor expansion \eqref{eq:Q-expansion} -- and thus of the potential term as expressed in \eqref{eq:pseudo-energy} -- is compromised. Although it does not mean that potential energy can diverge (it remains finite by definition), what it does mean is that such local linear approximations are generally challenged when the initial wave function has zeros.

\subsection{Geostrophic adjustment in the Madelung fluid and in shallow-water systems: discussion}

In light of the analogy between the Madelung fluid and two-dimensional rotating flows, it is insightful to compare the ageostrophic/non-adiabatic dynamics explored in \ref{part:adjustment} to the evolution of shallow-water flows in similar configurations. In such systems, if a perturbation occurs such that the flow is not in geostrophic balance, it then naturally relaxes toward a new geostrophic state by emitting away transient inertio-gravity (Poincaré) waves, which transport and redistribute energy and momentum while leaving potential vorticity unchanged. The theory behind this phenomenon, called geostrophic adjustment, was originally formulated by Rossby \cite{rossby1938mutual}, and developed further in the following decades \cite{cahn1945investigation,blumen1972geostrophic,schubert1980geostrophic,smith1981cyclostrophic,pedlosky2013geophysical,zeitlin2018geophysical}. The first experimental confirmations of the theory, by means of observations in oceanic currents, came only in the late seventies \cite{tang1979inertial,millot1981inertial}. In the atmosphere, observations of hurricanes such as \textit{Ivan} (2004) and \textit{Milton} (2024) provide graphical demonstrations of geostrophic adjustment: after the distributions of mass and potential vorticity in the centre of a cyclone change rapidly owing to deep convection bursts and latent heat release, the vortical flow subsequently adjusts to the new geostrophic balance by radiating gravity waves away from the storm's centre \cite{hendricks2010spontaneous,schubert2022tropical}. In two-dimensional flows, the phenomenon of geostrophic adjustment essentially relies on the conservation of potential vorticity and the fact that transient inertio-gravity waves do not transport it, whereas they can transport energy and momentum in order to adjust pressure and mass distribution to converge to a new geostrophically-balanced flow.

In contrast with geofluids\footnote{While reminiscent of the adjustment of geophysical flows, let us keep in mind that the oscillations of the Madelung fluid are a purely quantum phenomenon which should not be interpreted as a direct analogue of atmospheric dynamics}, the vorticity of the Madelung fluid is imposed by the cyclotron vector everywhere, which is a very constraining property that prevents the existence of similar propagation phenomena such as Poincaré and Rossby waves altogether, essential to the geostrophic adjustment of geophysical flows. Indeed, this constraint implies a profound difference in the restoring mechanism of the rotating Madelung fluid, as it constitutes an obstruction to Le Chatelier's principle. In the situation depicted throughout the paper, the response of the Madelung fluid is one-dimensional as the zonal velocity is enslaved to the instantaneous value of $b$ (Figure \ref{fig:mechanisms}). On the contrary, for a similar shallow-water flow, the generation of non-zero meridional velocity would produce a zonal Coriolis force and thus drive additional zonal velocity in the opposite direction as the original zonal flow, i.e. a counter-effect, which is essential to geostrophic adjustment. As a result, the persistent oscillations of frequency $2 b_1$ previously exhibited are in striking contrast with the adjustment of geostrophic flows where the final state is geostrophically balanced. In the geophysical context, there are possible obstructions to geostrophic adjustment, such as instabilities or the presence of boundaries where the transient inertio-gravity waves reflect and thus are unable to be emitted away from the flow. However, the Madelung fluid discussed throughout this paper is unbounded and a priori stable to any kind of perturbation, owing to the Hermiticity of the quantum system, yet it does not adjust to a new balanced state when $b$ reaches a new constant value, even if $b$ -- and thus the flow's vorticity -- is eventually returned to its initial value. As discussed in \cite{heifetz2023broglie}, the Madelung fluid locally supports de Broglie waves -- and the sloshing mode can be seen as a superposition of such waves --, but those cannot propagate away from a stationary flow that corresponds to a Landau level, since the Madelung fluid's density vanishes exponentially fast with $y^2$, which represents an effective harmonic trap for all waves. In other words, far from the density centre, waves cannot be emitted away, in contrast with transient gravity modes in open shallow-water systems. Besides, let us note that for the Madelung fluid, once the permanent regime is achieved, the persistent sloshing mode is a free oscillation, whereas the inertia-gravity waves observed being emitted away from the centre of a hurricane are the result of a varying source of mass or energy. One thus could, in principle, study the geostrophic adjustment of the rotating Madelung fluid in the presence of a similar source, and relate more closely the evolution of the corresponding quantum system to a hurricane-like phenomenon in the atmosphere.

\section{Conclusion and outlooks} \label{sec:conclusion}

The Madelung representation of the Schrödinger equation \eqref{eq:Schrodinger} describes a perfect compressible fluid whose dynamics obeys specific mechanisms that characterise the quantum nature of the wave function. Under a magnetic field, the Madelung fluid is subjected to a Lorentz force but its absolute vorticity remains strictly zero. On the other hand, the fluid is subjected to its Bohm potential, which is a function of the density that reflects the non-local nature of this fluid. These peculiar properties result in exotic hydrodynamical behaviour of the Madelung fluid, which can still be related to that of classical fluids, as far as particular situations are concerned. For instance, if the magnetic field varies slowly, the quasi-adiabatic evolution of the wave function shares similarities with the quasi-geostrophic adjustment of shallow-water flows.

The hydrodynamical perspective adopted in this paper allowed us to derive the exact solution to the initial quantum problem -- namely the sloshing mode -- without introducing quantum invariants as in \cite{lewis1969exact}, whereas the traditional perturbation theory presented in \ref{part:squeezing} only led to an approximate one that does not capture the self-similar nature of the wave function, nor the physical meaning of sloshing. Our derivation relied exclusively on the mechanistic interpretation of the Landau levels as stationary plane-Couette flows in geostrophic-like balance \cite{heifetz2021zero}, which allowed us to establish an analogy between ageostrophic dynamics and non-adiabatic quantum evolution. Beyond reproducing the known quantum dynamics, the Madelung representation revealed how departures from adiabaticity arise from the imbalance between magnetic confinement and quantum-pressure forces, leading to oscillatory energy exchange within the effective quantum fluid. We reported a natural connection between these mechanisms and the pseudo-energy, which quantifies the irreversibility of the quantum system in terms of ageostrophic dynamics of the fluid. We expect that this analogy goes beyond the context of this paper and can be insightful in other situations, especially since non-stationary quantum problems are ubiquitous in quantum physics and condensed matter literature. In this paper, we mainly focused on the evolution of a Landau level under a time-dependent magnetic field, however one can alternatively consider initially unbalanced states under a static magnetic field, for instance a wave packet on a lattice \cite{morais2024landau,razzoli2020continuous}, and develop the same Madelung framework for such cases. Other possible follow-up studies could address for instance i/ mean-field quantum interactions (with the Gross-Pitaevskii equation\footnote{Interestingly, while the linear Schrödinger equation yields non-linear Madelung equations, adding the non-linear term of the Gross-Pitaevskii equations results in a barotropic enthalpy contribution that is simply proportional to the density $\rho$.}) and their competition with the Bohm potential (local interactions against non-local quantum effects, healing length, etc.); ii/ boundaries in the meridional direction, which is expected to drastically modify the Landau level dynamics for Couette flows centred near those boundaries, and possibly shed light to the quantum Hall effect from the Madelung point of view; and iii/ perturbations in the zonal direction (thus two-dimensional) in order to explore de Broglie wave dynamics (which would be in the continuation of \cite{heifetz2023broglie}) or Berry phase effects \cite{berry1984quantal,xiao2010berry,sundaram1999wave}.

\funding{NP acknowledges support and funding from the international postdoctoral fellowship program of the Azrieli Foundation, and Tel Aviv University's postdoctoral program.}

\roles{N.P. wrote the article, derived the analytical expressions and performed the numerical simulations. Both authors discussed the physics, edited the text, gave final approval for publication and agreed to be held accountable for the work performed therein.}

\data{All data relating to the analytical results presented in this paper can be readily reproduced by a reader using the equations explicitly provided therein. The numerical calculations corresponding to the plots of Figures \ref{fig:b_VS_beth} and \ref{fig:reversibility} can be reproduced by the reader using the script provided at \url{https://github.com/NPerezTAU/Landau-level-sloshing}.}

\appendix

\section{Consistency between expressions (23) and (48)} \label{apx:match}

In section \ref{sec:fluid} we provide the exact solution \eqref{eq:wave_function_solution} of the Schrödinger equation \eqref{eq:Schrodinger_bis} with the initial condition \eqref{eq:IC_wave_function}, which must be consistent with the perturbative expression \eqref{eq:first-order-adiabatic} obtained in section \ref{sec:quantum}, when the approximation is valid. In the adiabatic limit, we always have
\begin{equation}
    |\Gamma(t) - b(t)| \ll b(t) \ ,
\end{equation}
thus we can expand the Hermite-Gauss function $\Psi_{n_0 , \Gamma}$ to the first order as
\begin{equation} \label{eq:Taylor_Gauss-Hermite}
    \begin{split}
        \Psi_{n_0 , \Gamma} (y) &\approx \Psi_{n_0 , b} (y) + (\Gamma -b) \left. \frac{\partial \Psi_{n_0 , \gamma}}{\partial \gamma} \right|_{\gamma = b} \\[6pt]
        &= \Psi_{n_0 , b} (y) + \left( \frac{\Gamma - b}{4b} \right) \frac{\ee^{-b y^2 / 2}}{\sqrt{2^n n!}} \left( \frac{b}{\pi} \right)^\frac{1}{4} \left[ H_n (y \sqrt{b}) - 2 b y^2 H_n (y \sqrt{b}) + 2 y \sqrt{b} H_n' (y \sqrt{b}) \right] \ ,
    \end{split}
\end{equation}
using definition \eqref{eq:Landau}. Then, using the two following properties of Hermite polynomials:
\begin{equation}
    \begin{split}
        H_n' (X) &= 2 n H_{n-1} (X) \\[6pt]
        H_{n+1} (X) &= 2 X H_n (X) - H_n' (X) \ ,
    \end{split}
\end{equation}
we can reduce the expansion \eqref{eq:Taylor_Gauss-Hermite} to
\begin{equation} \label{eq:Hermite_exp}
    \Psi_{n_0 , \Gamma} (y) \approx \Psi_{n_0 , b} (y) + \left( \frac{b - \Gamma}{4b} \right) \left[ \sqrt{(n_0 + 2)(n_0 + 1)} \Psi_{n_0 + 2 , b} (y) - \sqrt{n_0 (n_0 - 1)} \Psi_{n_0 - 2 , b} (y) \right] \ .
\end{equation}

Up to their respective phase factors -- $\exp \left( -\ii (n_0 + 1/2) \int_0^t b(t_1) \dd t_1 \right)$ for the perturbative expression \eqref{eq:first-order-adiabatic} and $\exp \left( \ii \beta(t) y^2 / 2 -\ii (n_0 + 1/2) \int_0^t \Gamma(t_1) \dd t_1 \right)$ for the exact solution \eqref{eq:wave_function_solution} -- both expressions are consistent as long as the $n_0 \pm 2$ projections match, which is true at short times since
\begin{equation}
    \begin{split}
        \ee^{-2 \ii \Theta(t)} \zeta(t) \ , \quad \ee^{2 \ii \Theta(t)} \zeta(t)^* \quad \text{and} \quad \frac{b(t) - \Gamma(t)}{4 b(t)}
    \end{split}
\end{equation}
are all equivalent to $\dot{b}(0) t / 4 b_0$ as $t \rightarrow 0^+$. Since $\beta(0)=0$, $\dot{\beta}(0)=0$ and $\dot{\Gamma}(0)=0$, the respective phase factor also match at short times and are approximately equal to $\exp \left( -\ii (n_0 + 1/2) b_0 t \right)$. As long as the evolution is adiabatic, i.e. $\Gamma(t) \approx b(t)$, the Taylor expansion \eqref{eq:Hermite_exp} holds even at long times. Conversely, one cannot ensure consistency between the phase factor and amplitude coefficients of the approximate expression \eqref{eq:first-order-adiabatic} and those of \eqref{eq:Hermite_exp}, derived from the exact solution \eqref{eq:wave_function_solution}, because the perturbative approach adopted in section \ref{sec:quantum} generally breaks down at long times.

\section{Solution of equation (49)} \label{apx:beta}

We derive here the unique solution of the non-linear differential equation \eqref{eq:beta_free} for the oscillations in the permanent regime (constant $b = b_1$). In order to do that, we define
\begin{equation} \label{eq:xi}
    \xi (t) = \exp \left( 2 \int_0^t \dd t' \, \beta(t') \right) \ .
\end{equation}

By successive time differentiation of \eqref{eq:xi}, we obtain
\begin{equation}
    \begin{split}
        \dot{\xi} &= 2\beta \xi \ , \\[6pt]
        \ddot{\xi} &= 2 \dot{\beta} \xi + 2 \beta \dot{\xi} = (2 \dot{\beta} + 4 \beta^2) \xi \ , \\[6pt]
        \dddot{\xi} &= 2(\ddot{\beta} + 6 \beta \dot{\beta} + 4 \beta^3) \xi \ ,
    \end{split}
\end{equation}
which, plugged into equation \eqref{eq:beta_free}, yields
\begin{equation} \label{eq:osc_xi}
    \dddot{\xi} = -8 b_1^2 \beta \xi = -4 b_1^2 \dot{\xi} \ .
\end{equation}

The general solution of \eqref{eq:osc_xi} is $\xi(t) = A + B \cos (2 b_1 t + \varphi)$ (note that the solution is valid regarding the definition \eqref{eq:xi} only if $|B/A| < 1$), therefore, defining $\epsilon = B/A$, we have
\begin{equation}
    \beta (t) = \frac{\dot{\xi}}{2 \xi} = \frac{- b_1 \epsilon \sin \left( 2 b_1 t + \varphi \right)}{1 + \epsilon \cos \left( 2 b_1 t + \varphi \right)} \ ,
\end{equation}
which corresponds to expression \eqref{eq:exact_solution}. Furthermore, $\Gamma (t)$ is obtained using the definition $\Gamma^2 = \dot{\beta} + \beta^2 + b^2$:
\begin{equation}
    \Gamma (t) = \frac{b_1 \sqrt{1 - \epsilon^2}}{1 + \epsilon \cos \left( 2 b_1 t + \varphi \right)} \ .
\end{equation}

Noticing from \eqref{eq:useful_relation} that $\xi = b_0 / \Gamma$, we can finally identify $A = b_0 / \left( b_1 \sqrt{1 - \epsilon^2} \right)$.

\section{Expression of energy integrals} \label{apx:proofs}

In this appendix we derive expressions \eqref{eq:injected_energy} and \eqref{eq:energy_partition}, which are based on the mathematical properties of Hermite polynomials. We use the bra-ket notation and the ladder operators of \ref{part:squeezing}, but define them for the trapping frequency $\Gamma$ instead of $b$ (this way the wave function derived in \ref{part:adjustment} is proportional to $| n_0 \rangle$ at any time, up to a phase factor that depends on $y$ and $t$). Thus, for the sloshing mode whose wave function is given by \eqref{eq:wave_function_solution}, we can write:
\begin{equation} \label{eq:demo1}
    \int \rho y^2 \, \dd y = \int |\psi|^2 y^2 \, \dd y = \langle n_0 | y^2 | n_0 \rangle = \frac{1}{2 \Gamma} \langle n_0 | (c + c^\dagger)^2 | n_0 \rangle = \frac{1}{2 \Gamma} \langle n_0 | ( c^2 + c^{\dagger 2} + c c^\dagger + c^\dagger c ) | n_0 \rangle \ .
\end{equation}

The terms coming from $c^2$ and $c^{\dagger 2}$ in \eqref{eq:demo1} vanish by definition of the action of the ladder operators. Moreover, since $c^\dagger c | n \rangle = n | n \rangle$ and the ladder operators obey the commutation rule $c c^\dagger - c^\dagger c = \text{Id}$, we finally have:
\begin{equation} \label{eq:y^2-beth}
    \int \rho y^2 \, \dd y = \frac{n_0 + 1/2}{\Gamma (t)} \ ,
\end{equation}
which proves expression \eqref{eq:injected_energy}. Similarly, the first two expressions of \eqref{eq:energy_partition}, i.e. the zonal and meridional kinetic energy of the flow $E_{kx}$ and $E_{ky}$, follow straightforwardly from relation \eqref{eq:y^2-beth}. As for the potential energy, using \eqref{eq:Bohm_beth} and \eqref{eq:density}, we obtain:
\begin{equation}
    E_Q = \int \rho Q \, \dd y = \alpha(t) \int \rho \, \dd y - \frac{\Gamma^2}{2} \int \rho y^2 \, \dd y = \left( n_0 + \frac{1}{2} \right) \frac{\Gamma(t)}{2} \ .
\end{equation}

\section{Numerical methods} \label{apx:numerics}

The numerical integrations whose outcomes are displayed in Figures \ref{fig:b_VS_beth} and \ref{fig:reversibility} are performed using the \texttt{Dedalus} solver (second version) \cite{burns2020dedalus}. The Python notebook -- in which we indicate the values of the parameters used in this paper -- can be found at \url{https://github.com/NPerezTAU/Landau-level-sloshing}. In this script, equation \eqref{eq:Schrodinger_bis} is numerically integrated using the implicit–explicit Runga–Kutta RK443 time stepper, and by projecting the wave function $\psi(y,t)$ on a basis of $N_y = 512$ Chebyshev polynomials, with a finite domain $y \in [-L_y / 2, L_y / 2]$ (we make sure that $L_y$ is large enough compared to the width of the wave function at all times, since the medium is normally unbounded) and the boundary conditions $\psi(y = -L_y / 2) = \psi(y = L_y / 2) = 0$. The function $b(t)$ used for Figures \ref{fig:b_VS_beth} and \ref{fig:reversibility} is a product of $\tanh$ functions. $\Gamma$ is computed at each time step using definition \eqref{eq:y^2-beth}, and the energies $E_Q, E_{kx}, E_{ky}$ are obtained using definition \eqref{eq:energy_def}.

\bibliographystyle{iopart-num}
\bibliography{IOP_bibliography}

\end{document}